# Four Tensors Determining the Thermal and Electric Conductivities of Non-Degenerate Electrons in Magnetized Plasma

M. V. Glushikhina*

*Space Research Institute, Russian Academy of Sciences, Moscow, 117997 Russia*

*\*e-mail: m.glushikhina@iki.rssi.ru*

**Abstract**—A solution to the Boltzmann equation is obtained for a magnetized plasma with non-degenerate electrons and ions by Chapman–Enskog method. To obtain an approximate solution, the Sonine polynomials up to the third-order approximation are used. Fully ionized plasma is considered. We obtained more accurately the components for the diffusion, thermal diffusion and diffusion thermoeffects tensors in comparison with previous publications on this subject.

## 1. INTRODUCTION

Heat and charge transfer in a magnetized nondegenerate plasma plays an important role in describing its behavior both in laboratory conditions and in the structure and evolution of stars. The kinetic coefficients such as thermal conductivity, diffusion, thermal diffusion and diffusion thermoeffect determine heat fluxes and current densities. Knowing the distributions of heat and current, we can calculate the magnetothermal evolution, the distribution of the magnetic field and temperature over the surface of stars, or describe the behavior of the plasma obtained and accelerated in laboratory conditions.

Classical methods of kinetic gas theory were developed by Maxwell, Boltzmann, Gilbert, Enskog, and Chapman. These methods are presented in the monograph by Chapman and Cowling [1]. They are based on the solution of the Boltzmann equation by the method of successive approximations. The thermodynamically equilibrium distribution function is taken as the zeroth approximation: for a nondegenerate gas, the Maxwell distribution; and if degeneracy is important, the Fermi–Dirac distribution. The equilibrium distribution function does not give an exact solution to the Boltzmann equation in the presence of inhomogeneity. Following [1], we look for a solution to the Boltzmann equation in the first approximation by expansion in the Sonine (Laguerre) polynomials. To take into account degeneracy, we use a system of orthogonal functions, which are a generalization of the Sonine polynomials, proposed in [2–4]; see also [5]. Usually, the first two terms of the expansion are taken to calculate the thermal conductivity. It was shown in [6] that such an approximation gives significant errors for the thermal conductivity coefficient, which become much smaller when the third-order polynomial expansion is used.

For the first time, the application of the Boltzmann equation to a gas of charged particles was made by Chapman [1]. Due to the divergence of the collision integral at large values of impact parameters for particles exhibiting Coulomb interaction, the average distance between particles was taken as the upper limit of integration in respect to the impact parameter. Thus, the viscosity, thermal conductivity, and diffusion coefficients were calculated for gases consisting of charged particles. The divergence of the collision integral for the Coulomb interaction with large impact parameters shows that the scattering of particles with a large impact parameter and a small change in momentum in a single collision plays a more important role than collisions with a large change in momentum. Landau used this fact to simplify the Boltzmann collision integral [7]. He expanded the distribution function after the collision with small changes in momentum and left the first two terms of the expansion.

The kinetic coefficients for a nondegenerate plasma were calculated in [8–11] in the presence and absence of a magnetic field using the Chapman–Enskog expansion method. Braginsky [12, 13] calculated the kinetic coefficients for a nondegenerate plasma in a magnetic field consisting of electrons and one sort of positively charged ions using kinetic equations normalized to different average velocities for ions and electrons. The Landau collision integral was used and two polynomials in the expansion were taken into account. The same approach was used in [14], which presented the calculation of kinetic coefficients for fully ionized plasma of complex composition. The kinetic coefficients for fully ionized plasma in a magnetic field were obtained by direct numerical calculation of the Fokker–Planck equation in [15]. The components of the thermal conductivity tensor for degenerate stellar nuclei were calculated in the Lorenz approximation for hydrogen plasma in [16] and [17].

A non-relativistic calculation, based on the quantum Lenard–Balescu transport equation for the thermal and electrical conductivities of plasma of highly degenerate, weakly coupled electrons, and nondegenerate, weakly coupled ions, was performed in [18]. The diffusion, thermal diffusion, and diffusion thermoeffect coefficients were calculated in [19] for plasma in a magnetic field with highly degenerate electrons and nondegenerate nuclei in the Lorentz approximation. In turn, the thermal conductivity tensor is calculated in [20] for arbitrarily degenerate electrons and nondegenerate nuclei in a magnetic field.

In the present work, we solve the Boltzmann equation by the Chapman–Enskog method for electrons in a nondegenerate plasma. The tensors of thermal diffusion, diffusion, and diffusion thermoeffect using the expansion of three polynomials are found, and the Lorentz gas example shows that the method has good convergence to the exact solution. An analytical expression is obtained for the components of these tensors in the three polynomial approximation, taking into account electron–electron collisions for the case of nondegenerate electrons in the presence of a magnetic field. Accounting for the third-order of the polynomial significantly improved the accuracy of the results. In the approximation of two polynomials, the obtained solution coincides with the published results.

## 2. BOLTZMANN AND TRANSPORT EQUATIONS

We use the Boltzmann equation for nondegenerate electrons in a magnetic field and take into consideration the interaction of electrons with ions and with one another. The Boltzmann equation, which describes the time variation of the electron distribution function $f$ in the presence of the electric and magnetic fields is written as [10, 11]

$$\frac{\partial f}{\partial t} + c_i \frac{\partial f}{\partial r_i} - \frac{e}{m_e}\left(E_i + \frac{1}{c}\varepsilon_{ikl}c_k B_l\right)\frac{\partial f}{\partial c_i} + J = 0. \quad (1)$$

Here, $(-e)$ and $m_e$ are the charge (negative) and the mass of the electron, $E_i$ and $B_i$ are the strength of the electric field and magnetic induction, $\varepsilon_{ikl}$ is the fully antisymmetric Levi-Civita tensor, and $c$ is the speed of the light. The collision integral $J$ for nondegenerate electrons and singly charged ions, from [1–4], is written in the form

$$J = J_{ee} + J_{eI} = \int [ff_1' - ff_1']g_{ee}bdbd\varepsilon dc_{1i} \\ + \int [ff_I' - ff_I']g_{eI}bdbd\varepsilon dc_{Ii}. \quad (2)$$

Here, the impact parameter $b$ and $\varepsilon$ are geometrical parameters of particle collisions with relative velocities $g_{ee}$ and $g_{eI}$.

The electron part of the collision integral in (2) is integrated over the phase space of the incoming particles $(dc_{1i})$ and the physical space of their arrival $(bdbd\varepsilon)$ [1]. The functions corresponding to velocities after collision are marked with primes.

The Boltzmann equation for electrons with the pair collision integral (2) can be applied in conditions when the electron gas is considered to be nearly ideal, i.e., the kinetic energy of electrons is much higher than the energy of electrostatic interactions. This condition is satisfied in the plasma of sufficiently low density. A detailed discussion of the applicability of the pair collision integral (2) and its modifications for nondegenerate high-density gases can be found in [1].

Let us introduce the thermal velocity of electrons, $v_i = c_i - c_{0i}$, where $c_{0i}$ is the mass-average velocity. Thus, we can write the Boltzmann equation with respect to the thermal velocity in the form [11]

$$\frac{df}{dt} + v_i\frac{\partial f}{\partial r_i} - \left[\frac{e}{m_e}\left(E_i + \frac{1}{c}\varepsilon_{ikl}v_k B_l\right) + \frac{dc_{0i}}{dt}\right]\frac{\partial f}{\partial v_i} \\ - \frac{e}{m_e c}\varepsilon_{ikl}v_k B_l\frac{\partial f}{\partial v_i} - \frac{\partial f}{\partial v_i}v_k\frac{\partial c_{0i}}{\partial r_k} + J = 0, \quad (3)$$

where

$$\frac{d}{dt} = \frac{\partial}{\partial t} + c_{0i}\frac{\partial}{\partial r_i}.$$

The transport equations for the electron concentration, total momentum, and energy of electrons in a two-component mixture of electrons and nuclei can be derived in the usual manner from the Boltzmann equation in a quasi-neutral plasma [1, 9–11] as

$$\frac{dn_e}{dt} + n_e\frac{\partial c_{0i}}{\partial r_i} + \frac{\partial}{\partial r_i}(n_e\langle v_i\rangle) = 0, \quad (4)$$

$$\rho\frac{dc_{0i}}{dt} = \frac{1}{c}\varepsilon_{ikl}j_k B_l, \quad (5)$$

$$\frac{3}{2}kn_e\frac{dT}{dt} - \frac{3}{2}kT\frac{\partial}{\partial r_i}(n_e\langle v_i\rangle) + \frac{\partial q_{ei}}{\partial r_i} \\ = j_i\left(E_i + \frac{1}{c}\varepsilon_{ikl}c_{0k}B_l\right) - \rho_e\langle v_i\rangle\frac{dc_{0i}}{dt}, \quad (6)$$

where

$$\langle v_{\alpha i}\rangle = \frac{1}{n_\alpha}\int f_\alpha v_{\alpha i}dc_{\alpha i}, \quad n_e = \int fdc_{ei}, \quad (7)$$

$$c_{0i} = \frac{1}{\rho}\sum_\alpha \rho_\alpha \langle c_{\alpha i}\rangle, \quad j_i = -n_e e \langle v_i\rangle, \tag{8}$$

$$q_{\alpha i} = \frac{1}{2} n_\alpha m_\alpha \langle v_\alpha^2 v_{\alpha i}\rangle, \quad \rho = \sum_\alpha m_\alpha n_\alpha. \tag{9}$$

Here, the summation is performed over the electrons and ions, $P_e = n_e m_e \langle v^2\rangle/3$, and when we neglect the electron viscosity, $P_e$ is the electron pressure; $\langle v_i\rangle$ is the average electron velocity in the comoving reference frame; $q_i$ is the electron heat flux; and $j_i$ is the electric current of electrons. Here and below, we assume the average mass velocity to be equal to the average ion velocity, $c_{0i} = \langle c_{Ii}\rangle$. We also take into account the electric current and heat flux produced only by electrons.

## 3. DERIVATION OF EQUATIONS FOR THE ELECTRON DISTRIBUTION FUNCTIONS IN THE FIRST APPROXIMATION

The Boltzmann equation can be solved by the Chapman–Enskog successive iteration method [1]. This method is used when the distribution functions are close to those in thermodynamic equilibrium, while the deviations from equilibrium are considered in the linear approximation. Equation for the second order deviation from the equilibrium distribution function was derived in [22] for a simple gas; see also [1]. The complexity of this equation, and rather narrow region where second order corrections could be important, strongly restricted the application of this approach.

The zeroth approximation to the electron distribution function is the Maxwell distribution, which is found by equating to zero the collision integral $J_{ee}$ from (2)

$$f_0 = n_e\left(\frac{m_e}{2\pi kT}\right)^{3/2}\exp^{(-m_e v^2/2kT)}, \quad \int f_0 dv_i = n_e. \tag{10}$$

Here, $k$ is Boltzmann's constant and $T$ is the temperature. The ion distribution function in the zeroth approximation $f_{I0}$ is assumed to be similar to the electron distribution function.

Using (10) in (4)–(9), we obtain the zeroth approximation for the transport equation; in this approximation $\langle v_i\rangle = 0$ and $q_i = 0$.

In the first approximation, we seek for the function $f$ in the form

$$f = f_0(1 + \chi). \tag{11}$$

The function $\chi$ admits representation of the solution in the form

$$\chi = -A_i\frac{\partial \ln T}{\partial r_i} - n_e D_i d_i, \tag{12}$$

$$d_i = \frac{\rho_N}{\rho}\frac{\partial \ln P_e}{\partial r_i} - \frac{\rho_e}{P_e}\frac{1}{\rho}\frac{\partial P_N}{\partial r_i} + \frac{e}{kT}\left(E_i + \frac{1}{c}\varepsilon_{ikl}c_{0k}B_l\right). \tag{13}$$

The plasma is supposed to be quasi-neutral. The functions $A_i$ and $D_i$ determine the heat transfer and diffusion, respectively. Substituting (12) in the equation for $\chi$ we obtain equations for $A_i$ and $D_i$ [1]. It was shown in [10, 11] that in the presence of a magnetic field with an axial vector $B_i$, the polar vectors $A_i$ and $D_i$ may also be sought in the form

$$\begin{aligned}A_i &= A^{(1)}v_i + A^{(2)}\varepsilon_{ijk}v_j B_k + A^{(3)}B_i(v_j B_j),\\ D_i &= D^{(1)}v_i + D^{(2)}\varepsilon_{ijk}v_j B_k + D^{(3)}B_i(v_j B_j),\end{aligned} \tag{14}$$

where $v_i$, $\epsilon_{ijk}v_j B_k$, $B_i(v_j B_j)$ are three linearly independent polar vectors and $A^{(\alpha)}$, $D^{(\alpha)}$, $\alpha = 1, 2, 3$ are functions of the scalars $v^2$ and $B^2$. Introducing functions

$$\xi_A = A^{(1)} + iBA^{(2)}, \quad \xi_D = D^{(1)} + iBD^{(2)}, \tag{15}$$

and dimensionless velocity $u_i\sqrt{m_e/(2kT)}v_i$, and omitting small (compared to unity) terms on the order of $m_e/m_I$, we obtain the equations for $\xi_A$ and $\xi_D$ in the form

$$\begin{aligned}&f_0\left(u^2 - \frac{5}{2}\right)u_i \\ &= -iBf_0\frac{e\xi_A}{m_e c}u_i + I_{ee}(\xi_A u_i) + I_{eI}(\xi_{AIi}u_{Ii}),\end{aligned} \tag{16}$$

$$f_0 u_i = -iBf_0\frac{e\xi_D}{m_e c}u_i + I_{ee}(\xi_D u_i) + I_{eI}(\xi_{DIi}u_{Ii}), \tag{17}$$

where

$$\begin{aligned}I_{ee}(\xi_k u_i) = \int f_0 f_{01}(\xi_k u_i + \xi_{k1}u_{1i} \\ - \xi'_k u'_i - \xi'_{k1}u'_{1i})g_{ee}bdbd\varepsilon dc_{1i},\end{aligned} \tag{18}$$

$$\begin{aligned}I_{eI}(\xi_k u_{Ii}) = \int f_0 f_{I0}(\xi_k u_i - \xi'_k u'_i)g_{eI}bdbd\varepsilon dc_{Ii},\\ k = A, D.\end{aligned} \tag{19}$$

According to [1], the solution for the functions $\xi_A$ and $\xi_D$ is sought in the form of a series of orthogonal polynomials. The Sonine polynomials are the expansion coefficients of the function $(1 - s)^{-\frac{5}{2}}\exp[xs/(1 - s)]$ in powers of $s$:

$$(1 - s)^{-\frac{5}{2}}\exp\left(\frac{xs}{1 - s}\right) = \Sigma S_{3/2}^{(p)}(x)s^p. \tag{20}$$

They are orthogonal,

$$\int_0^\infty e^{-x}S_{3/2}^{(p)}(x)S_{3/2}^{(q)}(x)x^{3/2}dx = \frac{\Gamma(p + 5/2)}{p!}\delta_{pq}, \tag{21}$$

and the first three polynomials are as follows:

$$S^{(0)}_{3/2}(x) = 1, \quad S^{(1)}_{3/2}(x) = \frac{5}{2} - x,$$
$$S^{(2)}_{3/2}(x) = \frac{35}{8} - \frac{7}{2}x + \frac{1}{2}x^2. \quad (22)$$

We are searching for $\xi_A$, $\xi_D$, $D^{(3)}$, and $A^{(3)}$ in the form

$$\xi_A = a_0 S^{(0)}_{3/2} + a_1 S^{(1)}_{3/2} + a_2 S^{(2)}_{3/2},$$
$$A^{(3)} = c_0 S^{(0)}_{3/2} + c_1 S^{(1)}_{3/2} + c_2 S^{(2)}_{3/2}, \quad (23)$$
$$\xi_D = d_0 S^{(0)}_{3/2} + d_1 S^{(1)}_{3/2} + d_2 S^{(2)}_{3/2},$$
$$D^{(3)} = z_0 S^{(0)}_{3/2} + z_1 S^{(1)}_{3/2} + z_2 S^{(2)}_{3/2}.$$

Multiplying (16) and (17) by $S^{(0)}_{3/2}(u^2)u_i$, $S^{(1)}_{3/2}(u^2)u_i$, and $S^{(2)}_{3/2}(u^2)u_i$, and integrating with respect to $dc_i$, we obtain systems of equations in a general form for the coefficients of thermal conductivity and thermal diffusion

$$\begin{cases} 0 = -\frac{3}{2}i\omega n_e a_0 + a_0(a_{00} + b_{00}) \\ + a_1(a_{01} + b_{01}) + a_2(a_{02} + b_{02}), \\ -\frac{15}{4}n_e = -\frac{15}{4}i\omega n_e a_1 + a_0(a_{10} + b_{10}) \\ + a_1(a_{11} + b_{11}) + a_2(a_{12} + b_{12}), \\ 0 = -\frac{105}{16}i\omega n_e a_2 + a_0(a_{20} + b_{20}) \\ + a_1(a_{21} + b_{21}) + a_2(a_{22} + b_{22}), \end{cases} \quad (24)$$

and for the diffusion coefficients and diffusion thermoeffect:

$$\begin{cases} -\frac{3}{2} = -\frac{3}{2}i\omega n_e d_0 + d_0(a_{00} + b_{00}) \\ + d_1(a_{01} + b_{01}) + d_2(a_{02} + b_{02}), \\ 0 = -\frac{15}{4}i\omega n_e d_1 + d_0(a_{10} + b_{10}) \\ + d_1(a_{11} + b_{11}) + d_2(a_{12} + b_{12}), \\ 0 = -\frac{105}{16}i\omega n_e d_2 + d_0(a_{20} + b_{20}) \\ + d_1(a_{21} + b_{21}) + d_2(a_{22} + b_{22}). \end{cases} \quad (25)$$

Here, $a_{jk}$ and $b_{jk}$ are matrix elements for collision integrals and $\omega = eB/(m_e c)$ is the cyclotron frequency.

## 4. MATRIX ELEMENTS $a_{jk}$ AND $b_{jk}$

To calculate the matrix elements, we introduce the following variables [1]:

$$G_{li} = \frac{1}{2}(c_i + c_{li}) = \frac{1}{2}(c'_i + c'_{li}),$$
$$g_{ee,i} = c_{li} - c_i, \quad g'_{ei,i} = c'_{li} - c'_i,$$

$$g_{ee} = |g_{ee,i}| = |g'_{ei,i}| = g'_{ee}, \quad G_{0i} = G_{li} - c_{0i}, \quad (26)$$
$$v_i = G_{0i} - \frac{1}{2}g_{ee,i}, \quad v_{i1} = G_{0i} + \frac{1}{2}g_{ee,i},$$
$$v^2 + v_1^2 = 2G_0^2 + \frac{1}{2}g_{ee}^2.$$

Here, $G_{li}$ is the velocity of the mass center of two colliding electrons in the laboratory reference system, $G_{0i}$ is the same value in the comoving reference system, $g_{ee,i}$ is the relative velocity of the colliding electrons before the collision, $g'_{ee,i}$ is the same value after the collision, and $v_i$ and $v_{1i}$ are the speeds of the colliding particles in the comoving reference frame defined above. We introduce dimensionless variables

$$g_i = \frac{1}{2}\left(\frac{m_e}{kT}\right)^{1/2} g_{ee,i}, \quad g'_i = \frac{1}{2}\left(\frac{m_e}{kT}\right)^{1/2} g'_{ee,i},$$
$$g = |g_i| = |g'_i| = g', \quad G_i = \left(\frac{m_e}{kT}\right)^{1/2} G_{0i},$$
$$dc_i dc_{1i} = \left(\frac{2kT}{m_e}\right)^3 dG_i dg_i, \quad g_i * g'_i = g^2 \cos(\theta), \quad (27)$$
$$u^2 + u_1^2 = G^2 + g^2, \quad u^2 = u_i^2,$$
$$u_1^2 = u_{1i}^2, \quad G^2 = G_i^2.$$

The elements

$$a_{j0} = \int f_0 f_{01} S^{(j)}_{3/2}(u^2) u_i [u_i + u_{1i} - u'_i - u'_{1i}] g b\, db\, d\varepsilon\, dg_i\, dG_i = 0 \quad (28)$$

are equal to zero, since the conservation of momentum in a collision nullifies the bracket in (28). The nonzero elements $a_{jk}$ ($j, k \geq 1$) are defined as

$$a_{jk} = \int f_0 f_{01} [S^{(j)}_{3/2}(u^2) u_i + S^{(j)}_{3/2}(u_1^2) u_{1i} \\ - S^{(j)}_{3/2}(u'^2) u'_i - S^{(j)}_{3/2}(u_1'^2) u'_{1i}] S^{(k)}_{3/2}(u^2) u_i g b\, db\, d\varepsilon\, dg_i\, dG_i. \quad (29)$$

In order to calculate the matrix elements $b_{jk}$ and $a_{jk}$, it is necessary, according to [1], to calculate the following typical integral:

$$\frac{1}{n_\alpha n_\beta} \int f_\alpha f_\beta [S^{(j)}_{3/2}(u^2) u_i \\ - S^{(j)}_{3/2}(u'^2) u'_i] S^{(k)}_{3/2}(u^2) u_i g_{\alpha\beta} b\, db\, d\varepsilon\, dc_{\alpha i}\, dc_{\beta i}, \quad (30)$$

where indices $\alpha$ and $\beta$ denote different particles.

Integrating as shown in [1], we can write

$$[S^{(j)}_{3/2}(u^2) u_i, S^{(k)}_{3/2}(u^2) u_i]_e \\ = \pi^{-3/2} \int e^{-g^2} \sum A_{jkrl} g^{2r} (1 - \cos^l(\theta)) g b\, db\, d\varepsilon\, dg, \quad (31)$$

where $A_{pqrl}$ is just a number, the formula for calculating which is presented in [1].

We introduce functions $\Omega_{ee}^{(l)}(r)$ as

$$\Omega_{ee}^{(l)}(r) = \sqrt{\pi} \int_0^\infty e^{-g^2} g^{2r+1} \phi_{ee}(l) dg, \quad (32)$$

where

$$\phi_{ee}(l) = \int_0^\infty (1 - \cos^l \theta) gb db. \quad (33)$$

Thus, (31) can be expressed in terms of (32):

$$[S_{3/2}^{(j)}(u^2)u_i, S_{3/2}^{(k)}(u^2)u_i]_e = 8 \sum A_{jkrl} \Omega_{ee}^{(l)}(r). \quad (34)$$

The matrix coefficients $a_{jk}$ are presented in the general form as

$$\begin{aligned} a_{11} &= 4 n_e^2 \Omega_{ee}^{(2)}(2), \\ a_{12} &= 7 n_e^2 \Omega_{ee}^{(2)}(2) - 2 n_e^2 \Omega_{ee}^{(2)}(3), \\ a_{22} &= \frac{77}{4} n_e^2 \Omega_{ee}^{(2)}(2) - 7 n_e^2 \Omega_{ee}^{(2)}(3) + n_e^2 \Omega_{ee}^{(2)}(4). \end{aligned} \quad (35)$$

The matrix elements $b_{jk}$ are calculated in a similar way; see [1].

We introduce the functions $\Omega_{ei}^{(l)}(r)$, similarly to (32):

$$\Omega_{ei}^{(l)}(r) = \sqrt{\pi} \int_0^\infty e^{-g^2} g^{2r+2} \phi_{ei}^{(l)} dg. \quad (36)$$

Here,

$$\phi_{ei}^{(l)} = \int_0^\infty (1 - \cos^l \theta) g_{ei} b db. \quad (37)$$

Integrating in (37) over the impact parameter $db$, we see that the integral logarithmically tends to infinity. It converges in a more accurate examination of Coulomb collisions in a plasma with allowance for the correlation functions [23], and an upper limit of integration $b_{max}$ arises.

For electron–ion collisions with $g_{ei} \sim v_e$ an approximate expression for the Coulomb logarithm is written as [24]

$$\Lambda = \ln\left(\frac{b_{max} \overline{v_e^2} m_e}{Ze^2}\right), \quad \Lambda \gg 1, \quad (38)$$

where

$$\overline{v_e^2} = \frac{3kT}{m_e}. \quad (39)$$

The expression for the electron–electron Coulomb logarithm is obtained from (38) for $Z = 1$. The value $b_{max}$ is the total radius of the Debye screening for electrons $r_{\mathcal{D}e}$ and ions $r_{\mathcal{D}i}$, which can be expressed as

$$\frac{1}{b_{max}^2} = \frac{1}{r_{\mathcal{D}i}^2} + \frac{1}{r_{\mathcal{D}e}^2} = \frac{4\pi e^2}{kT}(n_N Z^2 + n_e). \quad (40)$$

The average frequency of electron–ion collisions $\nu_{ei}$ in [25] in the limiting case of nondegenerate electrons is written as

$$\nu_{ei} = \frac{4}{3}\sqrt{\frac{2\pi}{m_e}} \frac{Z^2 e^4 n_N \Lambda}{(kT)^{3/2}}, \quad \tau_{nd} = 1/\nu_{nd}. \quad (41)$$

Using (36), the elements $b_{ij}$ of the symmetric matrix may be written in the following form:

$$b_{00} = 8 n_e n_i \Omega_{ei}^{(1)}(1), \quad (42)$$

$$b_{01} = 8 n_e n_i \left(\frac{5}{2} \Omega_{ei}^{(1)}(1) - \Omega_{ei}^{(1)}(2)\right), \quad (43)$$

$$b_{11} = 8 n_e n_i \left(\frac{25}{4} \Omega_{ei}^{(1)}(1) - 5 \Omega_{ei}^{(1)}(2) + \Omega_{ei}^{(1)}(3)\right), \quad (44)$$

$$b_{02} = 8 n_e n_i \left[\frac{35}{8} \Omega_{ei}^{(1)}(1) - \frac{7}{2} \Omega_{ei}^{(1)}(2) + \frac{1}{2} \Omega_{ei}^{(1)}(3)\right], \quad (45)$$

$$\begin{aligned} b_{12} = 8 n_e n_N &\left[\frac{175}{16} \Omega_{ei}^{(1)}(1) - \frac{105}{8} \Omega_{ei}^{(1)}(2) \right.\\ &\left. + \frac{19}{4} \Omega_{ei}^{(1)}(3) - \frac{1}{2} \Omega_{ei}^{(1)}(4)\right], \end{aligned} \quad (46)$$

$$\begin{aligned} b_{22} = 8 n_e n_i &\left[\frac{35^2}{8^2} \Omega_{ei}^{(1)}(1) - \frac{245}{8} \Omega_{ei}^{(1)}(2) \right.\\ &\left. + \frac{133}{8} \Omega_{ei}^{(1)}(3) - \frac{7}{2} \Omega_{ei}^{(1)}(4) + \frac{1}{4} \Omega_{ei}^{(1)}(5)\right]. \end{aligned} \quad (47)$$

According to [1], we find the expressions for $\Omega_{eN}^{(1)}(r)$ in the form

$$\Omega_{ei}^{(1)}(r) = \sqrt{\pi} \frac{e^4 \Lambda Z^2}{(2kT)^2} \Gamma(r), \quad \Gamma(1) = 1; \quad (48)$$
$$\Gamma(2) = 1; \quad \Gamma(3) = 2; \quad \Gamma(4) = 6; \quad \Gamma(5) = 24.$$

Using (41), we can write $b_{jk}$:

$$b_{00} = 8\sqrt{\pi} \frac{n_e n_N e^4 Z^2 \Lambda}{(2kT)^{3/2} \sqrt{m_e}} = \frac{3 n_e}{2\tau_{nd}}, \quad (49)$$

$$b_{01} = 12\sqrt{\pi} \frac{n_e n_N e^4 Z^2 \Lambda}{(2kT)^{3/2} \sqrt{m_e}} = \frac{9 n_e}{4\tau_{nd}}, \quad (50)$$

$$b_{11} = 26\sqrt{\pi} \frac{n_e n_N e^4 Z^2 \Lambda}{(2kT)^{3/2} \sqrt{m_e}} = \frac{39 n_e}{8\tau_{nd}}, \quad (51)$$

$$b_{02} = 15\sqrt{\pi} \frac{n_e n_N e^4 Z^2 \Lambda}{(2kT)^{3/2} \sqrt{m_e}} = \frac{45 n_e}{16\tau_{nd}}, \quad (52)$$

$$b_{12} = \frac{69\sqrt{\pi}}{2} \frac{n_e n_N e^4 Z^2 \Lambda}{(2kT)^{3/2} \sqrt{m_e}} = \frac{207 n_e}{32\tau_{nd}}, \quad (53)$$

$$b_{22} = \frac{433\sqrt{\pi}}{8} \frac{n_e n_N e^4 Z^2 \Lambda}{(2kT)^{3/2} \sqrt{m_e}} = \frac{1299 n_e}{128\tau_{nd}}. \quad (54)$$

For plasma with $\Lambda \gg 1$ from (38) we have

$$\phi_{ee}(2) \approx \frac{16 e^4}{m_e^2 g_{ee}^3}, \quad \Omega_{ee}^{(2)}(r) = \sqrt{\pi} \frac{e^4 \Lambda}{\sqrt{m_e}(kT)^{3/2}} \Gamma(r). \quad (55)$$

Using (55), we have from (35) with $n_e = Z n_i$:

$$a_{11} = 4 n_e^2 \frac{\sqrt{\pi} \Lambda_{ei} e^4}{\sqrt{m_e}(kT)^{3/2}} = \frac{3}{\sqrt{2}} \frac{n_e}{Z \tau_{nd}}, \quad (56)$$

$$a_{12} = 3 n_e^2 \frac{\sqrt{\pi} \Lambda_{ei} e^4}{\sqrt{m_e}(kT)^{3/2}} = \frac{9}{4\sqrt{2}} \frac{n_e}{Z \tau_{nd}}, \quad (57)$$

$$a_{22} = \frac{45}{4} n_e^2 \frac{\sqrt{\pi} \Lambda_{ei} e^4}{\sqrt{m_e}(kT)^{3/2}} = \frac{135}{16\sqrt{2}} \frac{n_e}{Z \tau_{nd}}. \quad (58)$$

## 5. EXPRESSIONS FOR TENSORS OF TRANSFER COEFFICIENTS IN A MAGNETIC FIELD

General expressions for the heat flux $q_i$ and average directional (diffusion) electron velocity $\langle v_i \rangle$ are given by

$$q_i = -\lambda_{ij} \frac{\partial T}{\partial x_j} - n_e v_{ij} d_j = q_i^{(A)} + q_i^{(D)}, \quad (59)$$

$$\langle v_i \rangle = -\mu_{ij} \frac{\partial T}{\partial x_j} - n_e \eta_{ij} d_j = \langle v_i^{(A)} \rangle + \langle v_i^{(D)} \rangle, \quad (60)$$

where $\lambda_{ij}$ and $v_{ij}$ are the thermal conductivity and diffusion thermoeffect tensors, respectively, while $\mu_{ij}$ and $\eta_{ij}$ are the thermal diffusion and diffusion tensors, respectively [5, 26]. The indices ($A$) and ($D$) correspond to the heat flux and diffusion velocity of the electrons determined by the temperature gradient $\partial T / \partial x_j$ and diffusion vector $d_j$, respectively. The transport coefficient tensors can be written as

$$\lambda_{ik} = \frac{5}{2} \frac{k^2 T n_e}{m_e} \Big\{ \big[ a_0^1 - a_1^1 \big] \delta_{ik} \\ - \varepsilon_{ikn} B_n \big[ b_0^1 - b_1^1 \big] + B_i B_k \big[ c_0^1 - c_1^1 \big] \Big\}, \quad (61)$$

$$v_{ik} = \frac{5}{2} \frac{k^2 T^2 n_e}{m_e} \Big\{ \big[ d_0^1 - d_1^1 \big] \delta_{ik} \\ - \varepsilon_{ikn} B_n \big[ e_0^1 - e_1^1 \big] + B_i B_k \big[ z_0^1 - z_1^1 \big] \Big\}, \quad (62)$$

$$\mu_{ik} = \frac{k}{m_e} \big\{ a_0^1 \delta_{ik} - \varepsilon_{ikn} B_n b_0^1 + B_i B_k c_0^1 \big\}, \quad (63)$$

$$\eta_{ik} = \frac{kT}{m_e} \big\{ d_0^1 \delta_{ik} - \varepsilon_{ikn} B_n e_0^1 + B_i B_k z_0^1 \big\}. \quad (64)$$

Here, $a_0^1$ and $b_0^1$ are the real and imaginary parts of coefficient $a_0$:

$$a_0 = a_0^1 + i B b_0^1, \quad a_1 = a_1^1 + i B b_1^1, \\ B^2 c_0^1 = (a_0^1)_{B=0} - a_0^1, \quad B^2 c_1^1 = (a_1^1)_{B=0} - a_1^1, \quad (65)$$

and $d_0^1$, $e_0^1$, $d_1^1$, and $e_1^1$ are the real and imaginary parts of coefficients $d_0$ and $d_1$:

$$d_0 = d_0^1 + i B e_0^1, \quad d_1 = d_1^1 + i B e_1^1, \\ B^2 z_0^1 = (d_0^1)_{B=0} - d_0^1, \quad B^2 z_1^1 = (d_1^1)_{B=0} - d_1^1. \quad (66)$$

The procedure of finding the components of the thermal conductivity tensor $\lambda_{ij}$ for arbitrary degeneracy was described in detail in [20], where analytic expressions for them were derived. For highly degenerate electrons, the coefficients of thermal diffusion, diffusion, and diffusion thermoeffect in the Lorentz approximation were derived in [19].

### 5.1. Thermal Diffusion Tensor for Nondegenerate Electrons

To obtain coefficient $a_0$, we must solve the system of equations (24) with the matrix elements $b_{jk}$ from (49)–(54) and the matrix elements $a_{jk}$ from (56)–(58). The system for a three-polynomial solution for electrons in the presence of a magnetic field, according to (24), taking into account (49)–(54) and (56)–(58), is written as

$$\begin{cases} 0 = -\frac{3}{2} i \omega \tau_{nd} a_0 + \frac{3}{2} a_0 + \frac{9}{4} a_1 + \frac{45}{16} a_2, \\ -\frac{15}{4} \tau_{nd} = -\frac{15}{4} i \omega \tau_{nd} a_1 + \frac{9}{4} a_0 \\ \quad + \frac{3}{2} \Big( \frac{13}{4} + \frac{\sqrt{2}}{Z} \Big) a_1 + \frac{9}{8} \Big( \frac{23}{4} + \frac{\sqrt{2}}{Z} \Big) a_2, \\ 0 = -\frac{105}{16} i \omega \tau_{nd} a_2 + \frac{45}{16} a_0 \\ \quad + \frac{9}{8} \Big( \frac{23}{4} + \frac{\sqrt{2}}{Z} \Big) a_1 + \frac{3}{32} \Big( \frac{433}{4} + \frac{45\sqrt{2}}{Z} \Big) a_2. \end{cases} \quad (67)$$

The first two equations for $a_2 = 0$ determine the two-polynomial approximation and give, taking (63) into account, the following results for the case $B = 0$:

$$a_0 = \frac{15}{4} \frac{\tau_{nd}}{1 + \sqrt{2}/Z}, \quad (68)$$

$$\mu_{nd}^{(2)} = \frac{15}{4} \frac{k}{m_e} \frac{\tau_{nd}}{1 + \sqrt{2}/Z} = 3.75 \frac{k}{m_e} \frac{\tau_{nd}}{1 + \sqrt{2}/Z}. \quad (69)$$

The results above are consistent with the results obtained in [9, 10].

In the three-polynomial approximation and for $B = 0$, we obtain solution (67) for $a_0$ in the form

$$a_0 = \frac{165}{32} \frac{1 + 15\sqrt{2}/(11Z)}{1 + 61\sqrt{2}/(16Z) + 9/(2Z^2)} \tau_{nd}, \quad (70)$$

$$\mu_{nd}^{(3)} = 5.1563 \frac{k}{m_e} \frac{1 + 15\sqrt{2}/(11Z)}{1 + 61\sqrt{2}/(16Z) + 9/(2Z^2)} \tau_{nd}. \quad (71)$$

The value

$$Q = \frac{32 m_e \mu_{nd}^{(3)}}{165 k \tau_{nd}} = \frac{1 + 15\sqrt{2}/(11Z)}{1 + 61\sqrt{2}/(16Z) + 9/(2Z^2)}, \quad (72)$$

which shows how much nondegenerate electron–electron collisions reduce the thermal diffusion coefficient at $B = 0$, is presented in Table 1 for various numbers $Z$.

In the two-polynomial approximation, taking into account the magnetic field and assuming $a_2 = 0$, we obtain a solution to system (67) in the form

$$a_0 = \frac{15}{4} \tau_{nd} \left[ 1 + \frac{\sqrt{2}}{Z} - \frac{5}{2} \omega^2 \tau_{nd}^2 - \left( \frac{23}{4} + \frac{\sqrt{2}}{Z} \right) i\omega\tau_{nd} \right]^{-1}, \quad (73)$$

$$a_0^1 = \frac{15}{4} \tau_{nd} \left( 1 + \frac{\sqrt{2}}{Z} - \frac{5}{2} \omega^2 \tau_{nd}^2 \right) \zeta, \quad (74)$$

$$b_0^1 = \frac{15}{4} \frac{\omega \tau_{nd}^2}{B} \left( \frac{23}{4} + \frac{\sqrt{2}}{Z} \right) \zeta, \quad (75)$$

where the coefficient

$$\zeta = \left[ \left( 1 + \frac{\sqrt{2}}{Z} \right)^2 + \left( \frac{449}{16} + \frac{13}{2} \frac{\sqrt{2}}{Z} + \frac{2}{Z^2} \right) \omega^2 \tau_{nd}^2 + \frac{25}{4} \omega^4 \tau_{nd}^4 \right]^{-1}. \quad (76)$$

In the three-polynomial approximation, the solution to system (67) has the form

$$a_0 = \frac{165}{32} \tau_{nd} \left( 1 + \frac{15\sqrt{2}}{11Z} - \frac{35}{11} i\omega\tau_{nd} \right) \frac{1}{\zeta_1 - i\omega\tau_{nd}\zeta_2}, \quad (77)$$

$$a_0^1 = \frac{165}{32} \tau_{nd} \left[ \left( 1 + \frac{15\sqrt{2}}{11Z} \right) \zeta_1 + \frac{35}{11} \omega^2 \tau_{nd}^2 \zeta_2 \right] \times \frac{1}{\zeta_1^2 + \omega^2 \tau_{nd}^2 \zeta_2^2}, \quad (78)$$

$$b_0^1 = \frac{165}{32} \frac{\omega \tau_{nd}^2}{B} \left[ -\frac{35}{11} \zeta_1 + \left( 1 + \frac{15\sqrt{2}}{11Z} \right) \zeta_2 \right] \times \frac{1}{\zeta_1^2 + \omega^2 \tau_{nd}^2 \zeta_2^2}, \quad (79)$$

where

$$\zeta_1 = 1 + \frac{61\sqrt{2}}{16Z} + \frac{9}{2Z^2} - \left( \frac{5385}{128} + \frac{365\sqrt{2}}{32Z} \right) \omega^2 \tau_{nd}^2, \quad (80)$$

**Table 1.** $Q$ values for various chemical elements: hydrogen ($Z = 26$), helium ($Z = 2$), carbon ($Z = 6$), oxygen ($Z = 8$), and iron ($Z = 26$) expected for the outer layers of white dwarfs and neutron stars

| $Z$ | 1 | 2 | 6 | 8 | 26 | $\infty$ |
|---|---|---|---|---|---|---|
| $Q$ | 0.268 | 0.407 | 0.653 | 0.712 | 0.885 | 1 |

$$\zeta_2 = \frac{1017}{64} + \frac{667\sqrt{2}}{32Z} + \frac{9}{2Z^2} - \frac{175}{16} \omega^2 \tau_{nd}^2. \quad (81)$$

The values of $c_0^1$ in the two and three-polynomial approximations are determined by using (65).

The diffusion velocity $\langle v_i^A \rangle$ using (60) and (63) can be written as

$$\langle v_i^A \rangle = -\frac{k}{m_e} \left[ a_0^1 \delta_{ik} - \varepsilon_{ikn} B_n b_0^1 + B_i B_k c_0^1 \right] \frac{\partial T}{\partial r_k} \quad (82)$$
$$= \langle v_i^A \rangle^{(1)} + \langle v_i^A \rangle^{(2)} + \langle v_i^A e \rangle^{(3)},$$

$$\langle v_i^A \rangle^{(1)} = -\frac{k}{m_e} a_0^1 \frac{\partial T}{\partial r_i} = -\mu_{nd}^{(1)} \frac{\partial T}{\partial r_i}, \quad (83)$$

$$\langle v_i^A \rangle^{(2)} = \frac{k}{m_e} \varepsilon_{ikn} B_n b_0^1 \frac{\partial T}{\partial r_k} = \varepsilon_{ikn} B_n \mu_{nd}^{(2)} \frac{\partial T}{\partial r_k}, \quad (84)$$

$$\langle v_i^A \rangle^{(3)} = -\frac{k}{m_e} B_i B_k c_0^1 \frac{\partial T}{\partial r_k} = -B_i B_k \mu_{nd}^{(3)} \frac{\partial T}{\partial r_k}. \quad (85)$$

For the two-polynomial approximation, we obtain

$$\mu_{nd}^{(12)} = \frac{k}{m_e} a_0^1 = \frac{15}{4} \frac{k}{m_e} \tau_{nd}$$
$$\times \frac{1 + \frac{\sqrt{2}}{Z} - \frac{5}{2} \omega^2 \tau_{nd}^2}{\left( 1 + \frac{\sqrt{2}}{Z} \right)^2 + \left( \frac{449}{16} + \frac{13}{2} \frac{\sqrt{2}}{Z} + \frac{2}{Z^2} \right) \omega^2 \tau_{nd}^2 + \frac{25}{4} \omega^4 \tau_{nd}^4}, \quad (86)$$

$$\mu_{nd}^{(22)} = -\frac{k}{m_e} b_0^1 = -\frac{15}{4} \frac{k}{m_e} \frac{\omega \tau_{nd}^2}{B}$$
$$\times \frac{\frac{23}{4} + \frac{\sqrt{2}}{Z}}{\left( 1 + \frac{\sqrt{2}}{Z} \right)^2 + \left( \frac{449}{16} + \frac{13}{2} \frac{\sqrt{2}}{Z} + \frac{2}{Z^2} \right) \omega^2 \tau_{nd}^2 + \frac{25}{4} \omega^4 \tau_{nd}^4}, \quad (87)$$

$$B^2 \mu_{nd}^{(32)} = \mu_{nd}^{(12)}(B = 0) - \mu_{nd}^{(12)}. \quad (88)$$

Expressions for thermal diffusion coefficients in the three-polynomial approximation can be written explicitly, using (78)–(85):

$$\mu_{nd}^{(13)} = \frac{k}{m_e} a_0^1$$
$$= \frac{165}{32} \frac{k}{m_e} \tau_{nd} \frac{\left( 1 + \frac{15\sqrt{2}}{11Z} \right) \zeta_1 + \frac{35}{11} \omega^2 \tau_{nd}^2 \zeta_2}{\zeta_1^2 + \omega^2 \tau^2 \zeta_2^2}, \quad (89)$$

$$\mu_{nd}^{(23)} = -\frac{k}{m_e} b_0^1$$

$$= -\frac{165}{32} \frac{k}{m_e} \frac{\omega \tau_{nd}^2}{B} \frac{-\frac{35}{11}\zeta_1 + \left(1 + \frac{15\sqrt{2}}{11Z}\right)\zeta_2}{\zeta_1^2 + \omega^2 \tau^2 \zeta_2^2}, \quad (90)$$

$$B^2 \mu_{nd}^{(33)} = \mu_{nd}^{(13)}(B = 0) - \mu_{nd}^{(13)}. \quad (91)$$

Based on (63) we can obtain a different form for writing the components of the thermal diffusion tensor in a magnetic field. Three components of the heat flux: parallel $\langle v_i^A \rangle_\parallel$, perpendicular $\langle v_i^A \rangle_\perp$ to the magnetic field **B**, and the Hall component $\langle v_i^A \rangle_{\text{hall}}$, perpendicular to both vectors $\nabla T$ and **B**, taking into account (69) or (71), are determined by the relations

$$\langle v_i^A \rangle_\parallel = -\mu_\parallel \nabla T_\parallel,$$

$$\mu_\parallel = \frac{k}{m_e}[a_0^1 + B^2 c_0^1] = \mu_{nd}, \quad (92)$$

$$\langle v_i^A \rangle_\perp = -\mu_\perp \nabla T_\perp, \quad \mu_\perp = \frac{k}{m_e} a_0^1, \quad (93)$$

$$\langle v_i^A \rangle_{\text{hall}} = -\mu_{\text{hall}} \frac{\nabla T \times \mathbf{B}}{B}, \quad \mu_{\text{hall}} = -\frac{k}{m_e} B b_0^1. \quad (94)$$

The difference between the two- and three-polynomial approximations can be characterized by comparing the values $Q_\perp^{(2)}$ and $Q_\perp^{(3)}$:

$$Q_\perp^{(2)} = \frac{\mu_{nd}^{(12)}}{\mu_{nd}^{(3)}}, \quad Q_\perp^{(3)} = \frac{\mu_{nd}^{(13)}}{\mu_{nd}^{(3)}}, \quad (95)$$

where $\mu_{nd}^{(12)}$ is defined in (86), $\mu_{nd}^{(3)}$ is defined in (71), and $\mu_{nd}^{(13)}$ is defined in (89). The functions $Q_\perp^{(2)}(\omega \tau_{nd})$ and $Q_\perp^{(3)}(\omega \tau_{nd})$ are presented in Fig. 1 for carbon, $Z = 6$. In this figure we have $Q_\perp^{(2)} = 0.3531$ and $Q_\perp^{(3)} = 0.1754$, when $\omega \tau = 0.25$.

### 5.2. Exact Solution in the Lorentz Approximation

The Lorentz approximation for solving the kinetic equation is used when the mass of light particles is much smaller than the mass of heavy particles, and, in addition, electron–electron collisions can be neglected. In this approximation, the linearized Boltzmann equation has an exact solution. This approximation works well for transport in a metal, where the strong electron degeneracy makes it possible to neglect electron–electron collisions. The Lorentz approximation can be used to verify the approximate polynomial solution, since it makes it possible to trace the convergence of the approximate solution to the exact one with increasing degree of polynomials. The solution in the Lorentz approximation was considered in different approaches [1, 16, 17, 27]; see also [24]. For thermal conductivity, the convergence of the polynomial solution to the exact one was considered in [20]. The exact explicit solution for the Lorentz approximation is obtained for the case $B = 0$:

$$\langle v_i^A \rangle = -\frac{128k}{\Lambda} \frac{m_e(kT)^3}{n_N n_e Z^2 e^4 h^3} \left( G_4 - \frac{5}{8} \frac{G_{5/2}}{G_{3/2}} G_3 \right) \frac{\partial T}{\partial r_i}. \quad (96)$$

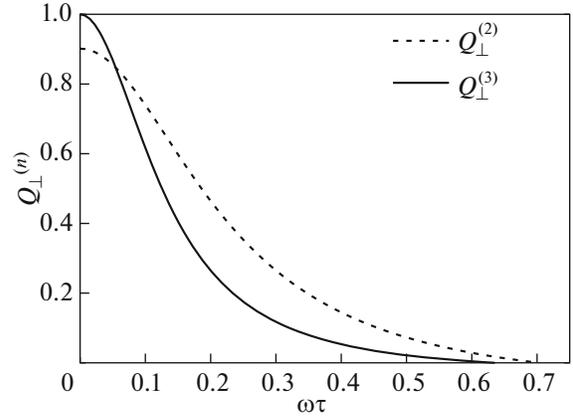

**Fig. 1.** Comparison of the two- and three-polynomial approximations for a nondegenerate carbon plasma with $Z = 6$ for different $\omega \tau$. The solid line shows the results for the three-polynomial approximation and the dashed line shows the results for the two-polynomial approximation.

Here, $G_i$ is the Fermi integral; see [24]. In limiting cases, the coefficient in (96) reduces to

$$\mu_e^l = \frac{12k}{\pi^{3/2} \Lambda n_N} \frac{(kT)^{3/2}}{e^4 Z^2 \sqrt{2m_e}} = \frac{16k}{m_e \pi} \tau_{nd}. \quad (97)$$

Exact formulas in the Lorentz model are used in [1] to estimate the accuracy of the polynomial approximation. The contribution of electron–electron collisions to the thermal diffusion coefficient for different $Z$ can be estimated from the plot of the normalized three-polynomial thermal diffusion coefficients in a direction perpendicular to the magnetic field by introducing quantity $Q_\perp^{(3l)}$ defined as

$$Q_\perp^{(3l)} = \frac{\mu_{nd}^{(13)}}{\mu_{e,nd}^l}. \quad (98)$$

Here, $\mu_{e,nd}^l$ is taken from the top line in (97). The curves for various $Z$, including $Z = \infty$, related to the Lorentz approximation, are shown in Fig. 2. The intersection of the curves with the $y$ axis in Fig. 2 occurs at the points given in Table 1, multiplied by $\mu_{nd}^{(3)}/\mu_{e,nd}^l = 1.0124$. When $\omega \tau = 0.25$ we have $Q_\perp^{(3l)} =$

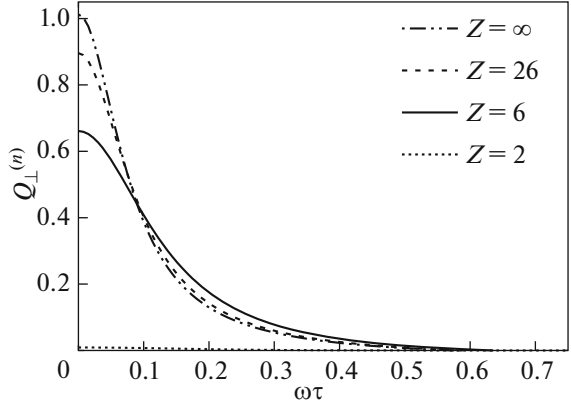

**Fig. 2.** The plots of values $Q_\perp^{(3l)}$ as a function $\omega\tau$ in the three-polynomial approximation for a nondegenerate helium plasma ($Z = 2$), carbon ($Z = 6$), iron ($Z = 26$) in comparison with the Lorentz plasma formally corresponding to $Z = \infty$. The deviation from the Lorentz plasma is due to the contribution of electron–electron collisions. The intersection of the Lorentz three-polynomial curve ($Z = \infty$) with the $Y$ axis at 1.0124 is associated with a deviation from the exact solution in the Lorentz approximation.

0.0824, 0.0908, 0.1159, and 0.0025 for $Z = \infty$, 26, 6, and 2, respectively.

### 5.3. Calculation of Polynomials without Taking into Account Electron–Electron Collisions, Thermal Diffusion Case

To estimate the accuracy of the polynomial approximation for the thermal diffusion coefficients, we compare them with the coefficients obtained as an exact solution in the Lorentz approximation. In the absence of a magnetic field in the Lorentz approximation with $a_{jk} = 0$, system (24) reduces to

$$\begin{cases} 0 = a_0 b_{00} + a_1 b_{01} + a_2 b_{02}, \\ -\dfrac{15}{4} n_e = a_0 b_{10} + a_1 b_{11} + a_2 b_{12}, \\ 0 = a_0 b_{20} + a_1 b_{21} + a_2 b_{22}. \end{cases} \quad (99)$$

Taking into account (49)–(54), we write this system as

$$\begin{cases} 0 = \dfrac{3}{2} a_0 + \dfrac{9}{4} a_1 + \dfrac{45}{16} a_2, \\ -\dfrac{15}{4} \tau_{nd} = \dfrac{9}{4} a_0 + \dfrac{39}{8} a_1 + \dfrac{207}{32} a_2, \\ 0 = \dfrac{45}{16} a_0 + \dfrac{207}{32} a_1 + \dfrac{1299}{128} a_2. \end{cases} \quad (100)$$

This system is written for the three-polynomial approximation. The first two equations for $a_2 = 0$ define a two-polynomial approximation, giving the following results, after taking into account (63):

$$a_0 = \dfrac{15}{4} \tau_{nd}, \quad \mu_{ndl}^{(2)} = \dfrac{15}{4} \dfrac{k}{m_e} \tau_{nd} = 3.75 \dfrac{k}{m_e} \tau_{nd}. \quad (101)$$

In the three-polynomial approximation, we obtain solution (100) for $a_0$ and, taking into account (63), the thermal diffusion coefficient in the form

$$a_0 = \dfrac{165}{32} \tau_{nd}, \quad \mu_{ndl}^{(3)} = \dfrac{165}{32} \dfrac{k}{m_e} \tau_{nd} = 5.1563 \dfrac{k}{m_e} \tau_{nd}. \quad (102)$$

The thermal diffusion obtained by the method of successive approximations of polynomials should be compared with the exact solution $\lambda_{nd}^l$ obtained by the Lorentz method (97) for nondegenerate electrons

$$\mu_{nd}^{(l)} = \dfrac{16}{\pi} \dfrac{k}{m_e} \tau_{nd} = 5.0931 \dfrac{k}{m_e} \tau_{nd}. \quad (103)$$

It is clear that the two-polynomial solution underestimates the coefficient of thermal diffusion by 26%, and the three-polynomial solution overestimates it by approximately 1.3%. The equations in the case of the three-polynomial approximation in the presence of a magnetic field were obtained from (24) with allowance for (49)–(54) and (100) in the form

$$\begin{cases} 0 = -\dfrac{3}{2} i\omega\tau_{nd} a_0 + \dfrac{3}{2} a_0 + \dfrac{9}{4} a_1 + \dfrac{45}{16} a_2, \\ -\dfrac{15}{4} \tau_{nd} = -\dfrac{15}{4} i\omega\tau_{nd} a_1 + \dfrac{9}{4} a_0 + \dfrac{39}{8} a_1 + \dfrac{207}{32} a_2, \\ 0 = -\dfrac{105}{16} i\omega\tau_{nd} a_2 + \dfrac{45}{16} a_0 + \dfrac{207}{32} a_1 + \dfrac{1299}{128} a_2. \end{cases} \quad (104)$$

The explicit solution of equations (104) for two- and three-polynomial approximations is determined by formulas (73)–(79) with the formally infinite value $Z$.

### 5.4. Partially Degenerate Electrons

For partially degenerate electrons, for which the matrix coefficients $b_{jk}$ were calculated in [20], with $x_0 = 0$ the level of degeneracy $DL = \varepsilon_{fe}/kT = 1.011$, system (24) is written in the form

$$\begin{cases} 0 = -1.5 i\omega n_e a_0 + a_0 b_{00} + a_1 b_{01} + a_2 b_{02}, \\ -3.88 n_e = -3.88 i\omega n_e a_1 + a_0 b_{10} + a_1 b_{11} + a_2 b_{12}, \\ 0 = -7.138 i\omega n_e a_2 + a_0 b_{20} + a_1 b_{21} + a_2 b_{22}. \end{cases} \quad (105)$$

In the absence of a magnetic field, this system reduces to

$$\begin{cases} 0 = 1.5 a_0 + 2.16 a_1 + 2.588 a_2, \\ -3.88 \tau_{d0} = 2.16 a_0 + 5.162 a_1 + 6.671 a_2, \\ 0 = 2.588 a_0 + 6.671 a_1 + 11.038 a_2. \end{cases} \quad (106)$$

**Table 2.** $U$ values for various chemical elements: hydrogen ($Z = 26$), helium ($Z = 2$), carbon ($Z = 6$), oxygen ($Z = 8$), and iron ($Z = 26$) expected for the outer layers of white dwarfs and neutron stars

| $Z$ | 1 | 2 | 6 | 8 | 26 | $\infty$ |
|---|---|---|---|---|---|---|
| $U$ | 0.575 | 0.684 | 0.836 | 0.868 | 0.949 | 1 |

The first two equations for $a_2 = 0$ determine the two-polynomial approximation, which, taking into account (63), gives the following result:

$$a_0 = 2.723\tau_{d0}, \quad \mu^{(2)}_{d0l} = 2.723 \frac{k}{m_e} \tau_{d0}. \quad (107)$$

In the three-polynomial approximation, taking into account (63), we obtain solution (106) for $a_0$, $a_1$ and also the thermal conductivity coefficient in the form

$$a_0 = 3.533\tau_{d0}, \quad \mu^{(3)}_{d0l} = 3.533 \frac{k}{m_e} \tau_{d0}. \quad (108)$$

Let us compare the thermal diffusion coefficient obtained by the method of successive approximations by polynomials with the exact solution $\mu^l_{nd}$ obtained by the Lorentz method (96) for nondegenerate electrons

$$\mu^{(l)}_{d0} = 0.744 \frac{16}{\pi} \frac{k}{m_e} \tau_{d0} = 3.789 \frac{k}{m_e} \tau_{d0}. \quad (109)$$

It can be seen that the two-polynomial solution underestimates the thermal diffusion coefficient by more than 28%, and the three-polynomial solution differs from the exact solution by approximately 7%.

## 6. TENSOR OF DIFFUSION AND DIFFUSION THERMOEFFECT FOR NONDEGENERATE ELECTRONS

For nondegenerate electrons, tensors (64) and (62) are written in the form

$$\eta_{ik} = \frac{kT}{m_e}\left[d_0^1 \delta_{ik} - \epsilon_{ijk} B_n e_0^1 + B_i B_k z_0^1 \right], \quad (110)$$

$$\nu_{ik} = \frac{5}{2} \frac{k^2 T^2}{m_e} n_e \left[ (d_0^1 - d_1^1)\delta_{ik} \right. \\ \left. - \epsilon_{ijk} B_n (e_0^1 - e_1^1) + B_i B_k (z_0^1 - z_1^1) \right]. \quad (111)$$

To find the coefficients $d_0$ and $d_1$ for arbitrary electron degeneration, similarly to the search for thermal diffusion coefficients, it is necessary to solve the system of equations (25) with the matrix elements $b_{jk}$ from (42)–(47) and the matrix elements $a_{jk}$.

For nondegenerate electrons, taking into account the three-polynomial expansion, system (25) should be written as

$$\begin{cases} -\dfrac{3}{2}\dfrac{\tau_{nd}}{n_e} = -\dfrac{3}{2}i\omega\tau_{nd}d_0 + \dfrac{3}{2}d_0 + \dfrac{9}{4}d_1 + \dfrac{45}{16}d_2, \\[4pt] 0 = -\dfrac{15}{4}i\omega\tau_{nd}d_1 + \dfrac{9}{4}d_0 + \dfrac{3}{2}\left(\dfrac{13}{4} + \dfrac{\sqrt{2}}{Z}\right)d_1 \\[4pt] \quad + \dfrac{9}{8}\left(\dfrac{23}{4} + \dfrac{\sqrt{2}}{Z}\right)d_2, \\[4pt] 0 = -\dfrac{105}{16}i\omega\tau_{nd}d_2 + \dfrac{45}{16}d_0 + \dfrac{9}{8}\left(\dfrac{23}{4} + \dfrac{\sqrt{2}}{Z}\right)d_1 \\[4pt] \quad + \dfrac{3}{32}\left(\dfrac{433}{4} + \dfrac{45\sqrt{2}}{Z}\right)d_2. \end{cases} \quad (112)$$

The first two equations for $d_2 = 0$ determine the two-polynomial approximation and give, taking into account (64) and (62), the following results for the case $B = 0$:

$$d_0 = -\frac{\tau_{nd}}{n_e} \frac{13/4 + \sqrt{2}/Z}{1 + \sqrt{2}/Z}, \quad d_1 = \frac{3}{2}\frac{\tau_{nd}}{n_e}\frac{1}{1 + \sqrt{2}/Z}, \quad (113)$$

$$\eta^{(2)}_{nd} = -\frac{kT}{m_e}\frac{\tau_{nd}}{n_e}\frac{13/4 + \sqrt{2}/Z}{1 + \sqrt{2}/Z}, \\ \nu^{(2)}_{nd} = -\frac{5}{2}\frac{k^2 T^2}{m_e}\tau_{nd}\frac{19/4 + \sqrt{2}/Z}{1 + \sqrt{2}/Z}. \quad (114)$$

The results above are consistent with the results obtained in [9, 10].

In the three-polynomial approximation and for $B = 0$, we obtain the solution (112) for $d_0$ and $d_1$, in the form

$$d_0 = -\frac{\tau_{nd}}{n_e}\frac{217/64 + 151\sqrt{2}/(16Z) + 9/(2Z^2)}{1 + 61\sqrt{2}/(16Z) + 9/(2Z^2)}, \quad (115)$$

$$d_1 = \frac{\tau_{nd}}{n_e}\frac{33/16 + 45\sqrt{2}/(16Z)}{1 + 61\sqrt{2}/(16Z) + 9/(2Z^2)}, \quad (116)$$

$$\eta^{(3)}_{nd} = -\frac{kT}{m_e}\frac{\tau_{nd}}{n_e}\frac{217/64 + 151\sqrt{2}/(16Z) + 9/(2Z^2)}{1 + 61\sqrt{2}/(16Z) + 9/(2Z^2)}, \quad (117)$$

$$\nu^{(3)}_{nd} = -\frac{5}{2}\frac{k^2 T^2}{m_e}\frac{349/64 + 196\sqrt{2}/(16Z) + 9/(2Z^2)}{1 + 61\sqrt{2}/(16Z) + 9/(2Z^2)}\tau_{nd}. \quad (118)$$

The values of

$$U = \frac{64 m_e n_e \eta^{(3)}_{nd}}{217 k T \tau_{nd}} \\ = -\frac{1 + 302\sqrt{2}/(17Z) + 288(217Z^2)}{1 + 61\sqrt{2}/(16Z) + 9/(2Z^2)}, \quad (119)$$

$$W = \frac{128 m_e \nu_{nd}^{(3)}}{1745 k^2 T^2 \tau_{nd}} \qquad (120)$$

$$= -\frac{1 + 784\sqrt{2}/(349Z) + 288/(349Z^2)}{1 + 61\sqrt{2}/(16Z) + 9/(2Z^2)},$$

which show the extent to which nondegenerate electron–electron collisions reduce the diffusion coefficient and diffusion thermoeffect coefficient at $B = 0$, are presented in Tables 2 and 3 for different values $Z$.

In the two-polynomial approximation, taking into account the magnetic field, assuming $d_2 = 0$, we obtain a solution to system (112) in the form

$$d_0 = \left(-\frac{13}{4} - \frac{\sqrt{2}}{Z} + \frac{5}{2} i \omega \tau_{nd}\right) \frac{\tau_{nd}}{n_e}$$

$$\times \left[1 + \frac{\sqrt{2}}{Z} - \frac{5}{2} \omega^2 \tau_{nd}^2 - \left(\frac{23}{4} + \frac{\sqrt{2}}{Z}\right) i \omega \tau_{nd}\right]^{-1}, \qquad (121)$$

$$d_0^1 = \frac{\tau_{nd}}{n_e}\left(-\frac{13}{4} - \frac{17\sqrt{2}}{4Z} - \frac{25}{4}\omega^2\tau_{nd}^2 - \frac{2}{Z^2}\right)\zeta, \qquad (122)$$

$$e_0^1 = -\frac{1}{n_e}\frac{\omega\tau_{nd}^2}{B}\left(\frac{259}{16} + \frac{13\sqrt{2}}{2Z} + \frac{2}{Z^2} + \frac{25}{4}\omega^2\tau_{nd}^2\right)\zeta, \qquad (123)$$

$$d_1 = \frac{13}{4}\frac{\tau_{nd}}{n_e}\left[1 + \frac{\sqrt{2}}{Z} - \frac{5}{2}\omega^2\tau_{nd}^2 - \left(\frac{23}{4} + \frac{\sqrt{2}}{Z}\right)i\omega\tau_{nd}\right]^{-1}, \qquad (124)$$

$$d_1^1 = \frac{\tau_{nd}}{n_e}\frac{3}{2}\left(1 + \frac{\sqrt{2}}{Z} - \frac{5}{2}\omega^2\tau_{nd}^2\right)\zeta, \qquad (125)$$

$$e_1^1 = \frac{1}{n_e}\frac{\omega\tau_{nd}^2}{B}\frac{3}{2}\left(\frac{23}{4} + \frac{\sqrt{2}}{Z}\right)\zeta, \qquad (126)$$

where the values of $\zeta$, $\zeta_1$, and $\zeta_2$ are defined in (76), (80), and (81).

In the three-polynomial approximation, the solution of system (112) has the form

$$d_0 = \frac{\tau_{nd}}{n_e}\left(-\frac{217}{64} - \frac{151\sqrt{2}}{16Z} - \frac{9}{2Z^2} + \frac{175}{16}\omega^2\tau_{nd}^2 + i\omega\tau_{nd}\left(\frac{3985}{128} + \frac{365\sqrt{2}}{32Z}\right)\right)\frac{1}{\zeta_1 - i\omega\tau_{nd}\zeta_2}, \qquad (127)$$

$$d_0^1 = \frac{\tau_{nd}}{n_e}\Bigg[\left(-\frac{217}{64} - \frac{151\sqrt{2}}{16Z} - \frac{9}{2Z^2} + \frac{175}{16}\omega^2\tau_{nd}^2\right)\zeta_1 - \omega^2\tau_{nd}^2\left(\frac{3985}{128} + \frac{365\sqrt{2}}{32Z}\right)\zeta_2\Bigg]\frac{1}{\zeta_1^2 + \omega^2\tau_{nd}^2\zeta_2^2}, \qquad (128)$$

$$e_0^1 = \frac{1}{n_e}\frac{\omega\tau_{nd}^2}{B}\Bigg[\left(\frac{3985}{128} + \frac{365\sqrt{2}}{32Z}\right)\zeta_1$$
$$-\left(\frac{217}{64} + \frac{151\sqrt{2}}{16Z} + \frac{9}{2Z^2} - \frac{175}{16}\omega^2\tau^2\right)\zeta_2\Bigg]\frac{1}{\zeta_1^2 + \omega^2\tau_{nd}^2\zeta_2^2}, \qquad (129)$$

$$d_1 = \frac{\tau_{nd}}{n_e}\left(\frac{33}{16} + \frac{45\sqrt{2}}{16Z} - \frac{105}{16}i\omega\tau_{nd}\right)\frac{1}{\zeta_1 - i\omega\tau_{nd}\zeta_2}, \qquad (130)$$

**Table 3.** $W$ values for various chemical elements: hydrogen ($Z = 26$), helium ($Z = 2$), carbon ($Z = 6$), oxygen ($Z = 8$), and iron ($Z = 26$) expected for the outer layers of white dwarfs and neutron stars

| $Z$ | 1 | 2 | 6 | 8 | 26 | $\infty$ |
|-----|-----|-----|-----|-----|-----|-----|
| $W$ | 0.459 | 0.579 | 0.767 | 0.809 | 0.925 | 1 |

$$d_1^1 = \frac{\tau_{nd}}{n_e}\Bigg[\left(\frac{33}{16} + \frac{45\sqrt{2}}{16Z}\right)\zeta_1 + \frac{105}{16}\omega^2\tau_{nd}^2\zeta_2\Bigg]$$
$$\times \frac{1}{\zeta_1^2 + \omega^2\tau_{nd}^2\zeta_2^2}, \qquad (131)$$

$$e_1^1 = \frac{1}{n_e}\frac{\omega\tau_{nd}^2}{B}\Bigg[-\frac{105}{16}\zeta_1 + \left(\frac{33}{16} + \frac{45\sqrt{2}}{16Z}\right)\zeta_2\Bigg]$$
$$\times \frac{1}{\zeta_1^2 + \omega^2\tau_{nd}^2\zeta_2^2}. \qquad (132)$$

The values of $z_0^1$ in the two- and three-polynomial approximations are determined using (66).

Speed $\langle v_i^D \rangle$ can be written in the form

$$\langle v_i^D \rangle = -n_e \frac{kT}{m_e}\Big[d_0^1 \delta_{ik} - \varepsilon_{ikn}B_n e_0^1 + B_i B_k z_0^1\Big]d_k$$
$$= \langle v_i^D \rangle^{(1)} + \langle v_i^D \rangle^{(2)} + \langle v_i^D \rangle^{(3)}. \qquad (133)$$

The heat flux $q_i^{(D)}$ can be written as

$$q_i^D = -n_e \frac{5}{2}\frac{k^2 T^2 n_e}{m_e}\Big[(d_0^1 - d_1^1)\delta_{ik}$$
$$-\varepsilon_{ikn}B_n(e_0^1 - e_1^1) + B_i B_k(z_0^1 - z_1^1)\Big]d_k \qquad (134)$$
$$= q_i^{D(1)} + q_i^{D(2)} + q_i^{D(3)}.$$

From this we get

$$\langle v_i^D \rangle^{(1)} = -n_e \frac{kT}{m_e}d_0^1 d_k = -n_e \eta_{nd}^{(1)} d_k, \qquad (135)$$

$$\langle v_i^D \rangle^{(2)} = n_e \frac{kT}{m_e}\varepsilon_{ikn}B_n e_0^1 d_k = n_e \varepsilon_{ikn} B_n \eta_{nd}^{(2)} d_k, \qquad (136)$$

$$\langle v_i^D \rangle^{(3)} = -n_e \frac{kT}{m_e}B_i B_k z_0^1 d_k = -n_e B_i B_k \eta_{nd}^{(3)} d_k, \qquad (137)$$

$$q_i^{D(1)} = -n_e \frac{k^2 T^2 n_e}{m_e}(d_0^1 - d_1^1)\delta_{ik}d_k = -n_e \nu_{nd}^{(1)} d_k, \qquad (138)$$

$$q_i^{D(2)} = n_e \frac{k^2 T^2 n_e}{m_e}\varepsilon_{ikn}B_n(e_0^1 - e_1^1)d_k$$
$$= n_e \varepsilon_{ikn}B_n \nu_{nd}^{(2)} d_k, \qquad (139)$$

$$q_i^{D(3)} = -n_e \frac{k^2 T^2 n_e}{m_e}B_i B_k(z_0^1 - z_1^1)d_k \qquad (140)$$
$$= -n_e B_i B_k \nu_{nd}^{(3)} d_k.$$

For the two-polynomial approximation, we obtain

$$\eta_{nd}^{(12)} = \frac{kT}{m_e} d_0^1 = \frac{kT}{m_e} \frac{\tau_{nd}}{n_e}$$

$$\times \frac{-\frac{13}{4} - \frac{17\sqrt{2}}{4Z} - \frac{2}{Z^2} - \frac{25}{4}\omega^2 \tau_{nd}^2}{\left(1 + \frac{\sqrt{2}}{Z}\right)^2 + \left(\frac{449}{16} + \frac{13}{2}\frac{\sqrt{2}}{Z} + \frac{2}{Z^2}\right)\omega^2 \tau_{nd}^2 + \frac{25}{4}\omega^4 \tau_{nd}^4}, \quad (141)$$

$$\eta_{nd}^{(22)} = -\frac{kT}{m_e} e_0^1 = -\frac{kT}{m_e} \frac{1}{n_e} \frac{\omega \tau_{nd}^2}{B}$$

$$\times \frac{-\frac{259}{16} - \frac{13\sqrt{2}}{2Z} - \frac{2}{Z^2} - \frac{25}{4}\omega^2 \tau_{nd}^2}{\left(1 + \frac{\sqrt{2}}{Z}\right)^2 + \left(\frac{449}{16} + \frac{13}{2}\frac{\sqrt{2}}{Z} + \frac{2}{Z^2}\right)\omega^2 \tau_{nd}^2 + \frac{25}{4}\omega^4 \tau_{nd}^4}, \quad (142)$$

$$B^2 \eta_{nd}^{(32)} = \eta_{nd}^{(12)}(B=0) - \eta_{nd}^{(12)}, \quad (143)$$

$$\nu_{nd}^{(12)} = \frac{5}{2}\frac{k^2T^2 n_e}{m_e}(d_0^1 - d_1^1) = \frac{5}{2}\frac{k^2T^2 n_e}{m_e}\frac{\tau_{nd}}{n_e}$$

$$\times \frac{-\frac{19}{4} - \frac{23\sqrt{2}}{4Z} - \frac{2}{Z^2} - \frac{5}{2}\omega^2 \tau_{nd}^2}{\left(1 + \frac{\sqrt{2}}{Z}\right)^2 + \left(\frac{449}{16} + \frac{13}{2}\frac{\sqrt{2}}{Z} + \frac{2}{Z^2}\right)\omega^2 \tau_{nd}^2 + \frac{25}{4}\omega^4 \tau_{nd}^4}, \quad (144)$$

$$\nu_{nd}^{(22)} = -\frac{5}{2}\frac{k^2T^2 n_e}{m_e}(e_0^1 - e_1^1) = -\frac{5}{2}\frac{k^2T^2 n_e}{m_e}\frac{1}{n_e}\frac{\omega \tau_{nd}^2}{B}$$

$$\times \frac{-\frac{397}{16} - \frac{8\sqrt{2}}{Z} - \frac{2}{Z^2} - \frac{25}{4}\omega^2 \tau_{nd}^2}{\left(1 + \frac{\sqrt{2}}{Z}\right)^2 + \left(\frac{449}{16} + \frac{13}{2}\frac{\sqrt{2}}{Z} + \frac{2}{Z^2}\right)\omega^2 \tau_{nd}^2 + \frac{25}{4}\omega^4 \tau_{nd}^4}, \quad (145)$$

$$B^2 \nu_{nd}^{(32)} = \nu_{nd}^{(12)}(B=0) - \nu_{nd}^{(12)}. \quad (146)$$

The expressions for diffusion coefficients in the three-polynomial approximation can be explicitly written using (128)–(132):

$$\eta_{nd}^{(13)} = \frac{kT}{m_e} d_0^1 = \frac{kT}{m_e}\frac{\tau_{nd}}{n_e} \frac{\left(-\frac{217}{64} - \frac{151\sqrt{2}}{16Z} - \frac{9}{2Z^2} + \frac{175}{16}\omega^2 \tau_{nd}^2\right)\zeta_1 - \omega^2 \tau^2 \left(\frac{3985}{128} + \frac{365\sqrt{2}}{32Z}\right)\zeta_2}{\zeta_1^2 + \omega^2 \tau^2 \zeta_2^2}, \quad (147)$$

$$\eta_{nd}^{(23)} = -\frac{kT}{m_e} e_0^1 = -\frac{kT}{m_e}\frac{\omega \tau_{nd}^2}{B n_e} \frac{\left(\frac{3985}{128} + \frac{365\sqrt{2}}{32Z}\right)\zeta_1 - \left(\frac{217}{64} + \frac{151\sqrt{2}}{16Z} + \frac{9}{2Z^2} - \frac{175}{16}\omega^2 \tau_{nd}^2\right)\zeta_2}{\zeta_1^2 + \omega^2 \tau^2 \zeta_2^2}, \quad (148)$$

$$B^2 \eta_{nd}^{(33)} = \eta_{nd}^{(13)}(B=0) - \eta_{nd}^{(13)}. \quad (149)$$

Similarly for the diffusion thermoeffect

$$\nu_{nd}^{(13)} = \frac{5}{2}\frac{k^2T^2 n_e}{m_e}(d_0^1 - d_1^1) = \frac{5}{2}\frac{k^2T^2 \tau_{nd}}{m_e}\frac{\left(-\frac{349}{64} - \frac{49\sqrt{2}}{4Z} - \frac{9}{2Z^2} + \frac{175}{16}\omega^2 \tau_{nd}^2\right)\zeta_1 - \omega^2 \tau^2 \left(\frac{4825}{128} + \frac{365\sqrt{2}}{32Z}\right)\zeta_2}{\zeta_1^2 + \omega^2 \tau^2 \zeta_2^2}, \quad (150)$$

$$\nu_{nd}^{(23)} = -\frac{5}{2}\frac{k^2T^2 n_e}{m_e}(e_0^1 - e_1^1) = -\frac{5}{2}\frac{k^2T^2}{m_e}\frac{\omega \tau_{nd}^2}{B}\frac{\left(\frac{4825}{128} + \frac{365\sqrt{2}}{32Z}\right)\zeta_1 - \zeta_2\left(\frac{349}{64} + \frac{196\sqrt{2}}{16Z} + \frac{9}{2Z^2} - \frac{175}{16}\omega^2 \tau_{nd}^2\right)}{\zeta_1^2 + \omega^2 \tau^2 \zeta_2^2}, \quad (151)$$

$$B^2 \nu_{nd}^{(33)} = \nu_{nd}^{(13)}(B=0) - \nu_{nd}^{(13)}. \quad (152)$$

Further, we obtain a different form for presenting the components of the diffusion tensors and the diffusion thermoeffect in a magnetic field similarly to the thermal diffusion case. Three components of the heat flux and diffusion velocity—parallel $\langle v_i^D \rangle_\parallel$, $q_\parallel^D$ to the magnetic field **B**; perpendicular $\langle v_i^D \rangle_\perp$, $q_\perp^D$ to the magnetic field **B**; and Hall components $\langle v_i^D \rangle_{hall}$, $q_{hall}^D$ perpendicular to both vectors $\nabla T$ and **B**—are determined by the relations

$$\langle v_i^D \rangle_\parallel = -n_e \eta_\parallel d_\parallel, \quad q_\parallel^D = -n_e \nu_\parallel d_\parallel, \quad (153)$$

$$\eta_\parallel = \frac{kT}{m_e}[d_0^1 + B^2(z_0^1)] = \eta_{nd}, \quad (154)$$

$$\nu_\parallel = \frac{5}{2}\frac{k^2T^2 n_e}{m_e}[(d_0^1 - d_1^1) + B^2(z_0^1 - z_1^1)] = \nu_{nd}, \quad (155)$$

$$\langle v_i^D \rangle_\perp = -n_e \eta_\perp d_\perp, \quad \eta_\perp = \frac{kT}{m_e}(d_0^1), \quad (156)$$

$$q_\perp^D = -n_e \nu_\perp d_\perp, \quad \nu_\perp = \frac{5}{2}\frac{k^2T^2 n_e}{m_e}(d_0^1 - d_1^1), \quad (157)$$

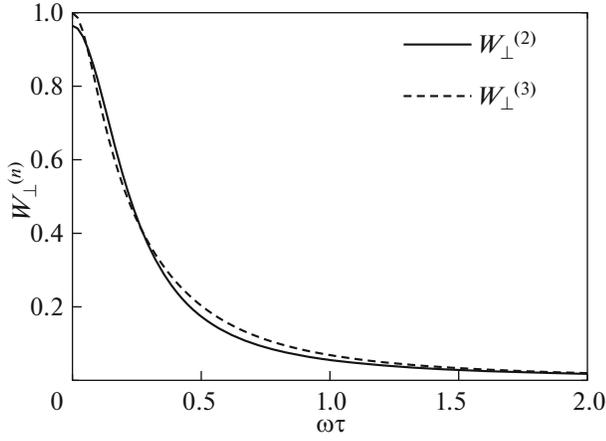

**Fig. 3.** Diffusion thermoeffect. Comparison of the two- and three-polynomial approximations for nondegenerate carbon plasma with $Z = 6$ for different $\omega\tau$.

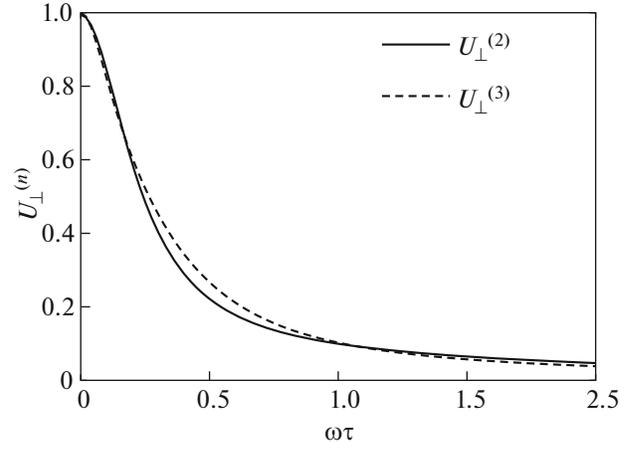

**Fig. 4.** Diffusion. Comparison of the two- and three-polynomial approximations for nondegenerate carbon plasma with $Z = 6$ for different $\omega\tau$.

$$\langle v_i^D \rangle_{hall} = -n_e \eta_{hall} \frac{\mathbf{d} \times \mathbf{B}}{B}, \quad \eta_{hall} = \frac{kT}{m_e} B e_0^1. \quad (158)$$

$$q_{hall}^D = -n_e \nu_{hall} \frac{\nabla T \times \mathbf{B}}{B},$$
$$\nu_{hall} = \frac{5}{2} \frac{k^2 T^2 n_e}{m_e} B(e_0^1 - e_0^1). \quad (159)$$

The results for the two polynomials coincide with the corresponding results obtained in [9, 10].

The difference between two- and three-polynomial approximations can be characterized by comparing the values of $U_\perp^{(2)}$, $W_\perp^{(2)}$ and $U_\perp^{(3)}$, $W_\perp^{(3)}$:

$$U_\perp^{(2)} = \frac{\eta_{nd}^{(12)}}{\eta_{nd}^{(3)}}, \quad U_\perp^{(3)} = \frac{\eta_{nd}^{(13)}}{\eta_{nd}^{(3)}},$$
$$W_\perp^{(2)} = \frac{\nu_{nd}^{(12)}}{\nu_{nd}^{(3)}}, \quad W_\perp^{(3)} = \frac{\nu_{nd}^{(13)}}{\nu_{nd}^{(3)}}, \quad (160)$$

where $\eta_{nd}^{(12)}$, $\nu_{nd}^{(12)}$ are defined in (141), (144), $\eta_{nd}^{(3)}$, $\nu_{nd}^{(3)}$ are defined in (118), (117), and $\eta_{nd}^{(13)}$, $\nu_{nd}^{(13)}$ are defined in (147), (150). The functions $U_\perp^{(2)}(\omega\tau_{nd})$, $W_\perp^{(2)}(\omega\tau_{nd})$, $U_\perp^{(3)}(\omega\tau_{nd})$, and $W_\perp^{(3)}(\omega\tau_{nd})$ are presented in Figs. 3–5 for carbon, $Z = 6$.

The system of equations (4)–(6) can be supplemented by Maxwell's equations

$$\nabla \times \mathbf{E} = -\frac{1}{c} \frac{\partial \mathbf{B}}{\partial t}, \quad \nabla \cdot \mathbf{B} = 0, \quad \nabla \times \mathbf{B} = \frac{4\pi}{c} \mathbf{j}.$$

For scalar conductivity $\sigma$, when

$$\mathbf{j} = \sigma \left( \mathbf{E} + \frac{1}{c} \mathbf{v} \times \mathbf{B} \right), \quad (161)$$

the equation for the magnetic field has the form [28]

$$\frac{\partial \mathbf{B}}{\partial t} = \nabla \times (\mathbf{v} \times \mathbf{B}) + \frac{c^2}{4\pi\sigma} \Delta \mathbf{B}. \quad (162)$$

The expression for the electric current vector $j_i$ is more complex when we consider the strict form of kinetic coefficients:

$$j_i = -en_e \langle v_i \rangle = -en_e \left[ \langle v_i^{(A)} \rangle + \langle v_i^{(D)} \rangle \right]$$
$$= -en_e \left[ -\mu_{ij} \frac{\partial T}{\partial x_j} - n_e \eta_{ij} d_j \right]. \quad (163)$$

If we present components $j_i$ with respect to the direction of the magnetic field $\mathbf{B}$, the electric current of electrons in the plasma can be written as

$$j_\| = en_e(n_e\eta_\| d_\| + \mu_\| \nabla T_\|),$$
$$j_\perp = en_e(n_e\eta_\perp d_\perp + \mu_\perp \nabla T_\perp), \quad (164)$$
$$j_{\text{hall}} = en_e\left(n_e\eta_{\text{hall}}\frac{\mathbf{d}\times\mathbf{B}}{B} + \mu_{\text{hall}}\frac{\nabla T\times\mathbf{B}}{B}\right).$$

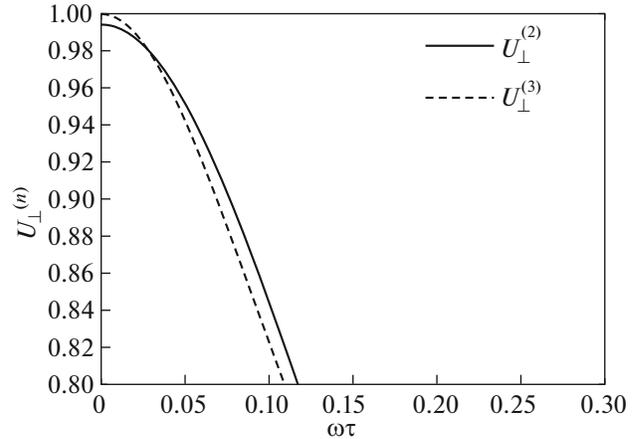

**Fig. 5.** An enlarged segment of the plot comparing the two- and three-polynomial approximations for diffusion with $\omega\tau < 0.15$.

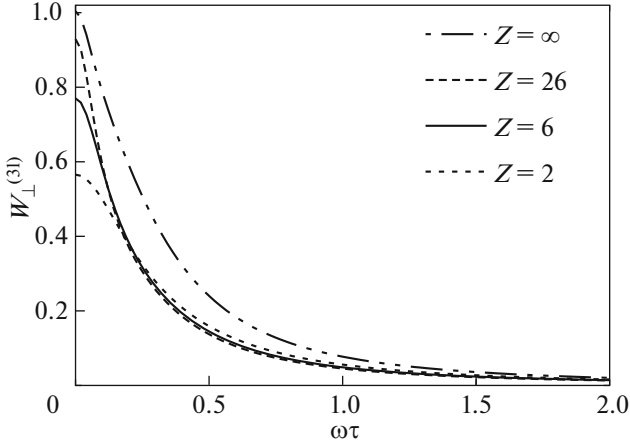

**Fig. 6.** The plots of the values $W_\perp^{(3l)}$ of the diffusion thermoeffect as a function $\omega\tau$ in the three-polynomial approximation for a nondegenerate plasma of helium ($Z = 2$), carbon ($Z = 6$), and iron ($Z = 26$) in comparison with the Lorentz plasma formally corresponding to $Z = \infty$. The intersection of the Lorentz three-polynomial curve ($Z = \infty$) with the Y axis at 1.0038 is associated with a deviation from the exact solution in the Lorentz approximation.

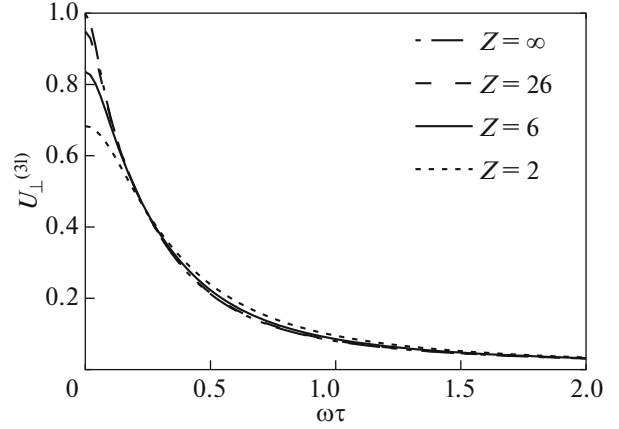

**Fig. 7.** The plots of the diffusion values $U_\perp^{(3l)}$ as a function $\omega\tau$ in the three-polynomial approximation for a nondegenerate plasma of helium ($Z = 2$), carbon ($Z = 6$), and iron ($Z = 26$) in comparison with the Lorentz plasma formally corresponding to $Z = \infty$. The intersection of the Lorentz three-polynomial curve ($Z = \infty$) with the Y axis at 0.9985 is associated with a deviation from the exact solution in the Lorentz approximation.

For completeness, we can write the expressions for the thermal conductivity tensor obtained for the three-polynomial approximation in [20]:

$$\lambda_{nd}^{(13)} = \frac{5}{2}\frac{k^2 T n_e}{m_e}(a_0^1 - a_1^1) = \frac{2125}{64}\frac{k^2 T n_e}{m_e}\tau_{nd}$$
$$\times\frac{\left(1 + \frac{18\sqrt{2}}{17Z} - \frac{14}{17}\omega^2\tau_{nd}^2\right)\zeta_1 + \left(\frac{1133}{340} + \frac{9\sqrt{2}}{17Z}\right)\omega^2\tau_{nd}^2\zeta_2}{\zeta_1^2 + \omega^2\tau^2\zeta_2^2}, \quad (165)$$

$$\lambda_{nd}^{(23)} = -\frac{5}{2}\frac{k^2 T n_e}{m_e}(b_0^1 - b_1^1) = -\frac{2125}{64}\frac{k^2 T n_e}{m_e}\frac{\omega\tau_{nd}^2}{B}$$
$$\times\frac{\left(1 + \frac{18\sqrt{2}}{17Z} - \frac{14}{17}\omega^2\tau^2\right)\zeta_2 - \left(\frac{1133}{340} + \frac{9\sqrt{2}}{17Z}\right)\zeta_1}{\zeta_1^2 + \omega^2\tau^2\zeta_2^2}, \quad (166)$$

$$B^2\lambda_{nd}^{(33)} = \lambda_{nd}^{(13)}(B=0) - \lambda_{nd}^{(13)}. \quad (167)$$

### 6.1. Exact Solution in the Lorentz Approximation for Diffusion and Diffusion Thermoeffect

Similarly to Section 5.2, we obtain the exact solution for the Lorentz approximation for $B = 0$:

$$\langle v_i^D\rangle = -n_e\frac{G_{5/2}}{G_{3/2}}\eta_{ij}d_j = \frac{32}{\Lambda}\frac{m_e(kT)^4}{n_e n_N Z^2 e^4 h^3}\frac{G_{5/2}}{G_{3/2}}G_3 d_i, \quad (168)$$

$$q_i^D = -n_e\frac{G_{5/2}}{G_{3/2}}\nu_{ij}d_j = \frac{128}{\Lambda}\frac{m_e(kT)^5}{n_N Z^2 e^4 h^3}\frac{G_{5/2}}{G_{3/2}}G_4\frac{\partial T}{\partial r_i}. \quad (169)$$

In the limiting cases, the coefficients in (168), (169) are reduced to

$$\eta_e^l = -\frac{32}{3\pi}\frac{kT}{n_e m_e \pi}\tau_{nd}, \quad (170)$$

$$\nu_e^l = -\frac{128(kT)^2}{3\pi m_e}\tau_{nd}. \quad (171)$$

The contribution of electron–electron collisions to the diffusion and diffusion thermoeffect coefficients for different $Z$ can be estimated from the plots for normalized three-polynomial coefficients in the direction perpendicular to the magnetic field, introducing the values $U_\perp^{(3l)}$ and $W_\perp^{(3l)}$, defined as

$$U_\perp^{(3l)} = \frac{\eta_{nd}^{(13)}}{\eta_e^l}, \quad W_\perp^{(3l)} = \frac{\nu_{nd}^{(13)}}{\nu_e^l}. \quad (172)$$

The curves for various $Z$, including $Z = \infty$ related to the Lorentz approximation, are shown in Figs. 6 and 7. The intersection of the curves with the y axis in the figures occurs at the points given in Tables 2 and 3 multiplied by $\eta_{nd}^{(3)}/\eta_e^l = 0.9985$ and $\nu_{nd}^{(3)}/\nu_e^l = 1.0038$. If $\omega\tau = 1$, we have $U_\perp^{(3l)} = 0.0079$, 0.081, 0.0086, and 0.095 for $Z = \infty$, 26, 6, 2, respectively, and $W_\perp^{(3l)} = 0.0443$, 0.0455, 0.0488, and 0.0563 for $Z = \infty$, 26, 6, and 2.

### 6.2. Calculation of Polynomials without Taking into Account Collisions between Electrons

To check the accuracy of the polynomial approximation for the diffusion coefficients and the diffusion thermoeffect, we compare them with the coefficients obtained as an exact solution in the Lorentz approximation. In the absence of a magnetic field, with $a_{jk} = 0$, system (112) reduces to

$$\begin{cases} -\dfrac{3}{2} = d_0 b_{00} + d_1 b_{01} + d_2 b_{02}, \\ 0 = d_0 b_{10} + d_1 b_{11} + d_2 b_{12}, \\ 0 = d_0 b_{20} + d_1 b_{21} + d_2 b_{22}. \end{cases} \quad (173)$$

Taking into account (49)–(54), we write this system as

$$\begin{cases} -\dfrac{3}{2}\dfrac{\tau_{nd}}{n_e} = \dfrac{3}{2} d_0 + \dfrac{9}{4} d_1 + \dfrac{45}{16} d_2, \\ 0 = \dfrac{9}{4} d_0 + \dfrac{39}{8} d_1 + \dfrac{207}{32} d_2, \\ 0 = \dfrac{45}{16} d_0 + \dfrac{207}{32} d_1 + \dfrac{1299}{128} d_2. \end{cases} \quad (174)$$

This system is written for the three-polynomial approximation. The first two equations with $d_2 = 0$ determine the two-polynomial approximation, giving the following results:

$$d_0 = -\dfrac{13}{4}\dfrac{\tau_{nd}}{n_e}, \quad d_1 = \dfrac{3}{2}\dfrac{\tau_{nd}}{n_e},$$

$$\eta_{ndl}^{(2)} = -\dfrac{13}{4}\dfrac{kT}{m_e}\dfrac{\tau_{nd}}{n_e} = -3.25\dfrac{kT}{m_e}\dfrac{\tau_{nd}}{n_e}, \quad (175)$$

$$\nu_{ndl}^{(2)} = -\dfrac{5}{2}\dfrac{k^2 T^2 n_e}{m_e}\dfrac{\tau_{nd}}{n_e}\dfrac{19}{4} = -11.875\dfrac{k^2 T^2}{m_e}\tau_{nd}.$$

In the three-polynomial approximation, we obtain solution (174) for $d_0$, $d_1$, and the coefficients in the form

$$d_0 = -\dfrac{217}{64}\dfrac{\tau_{nd}}{n_e}, \quad d_1 = \dfrac{33}{16}\dfrac{\tau_{nd}}{n_e},$$

$$\eta_{ndl}^{(3)} = -\dfrac{217}{64}\dfrac{kT}{m_e}\dfrac{\tau_{nd}}{n_e} = -3.3906\dfrac{kT}{m_e}\dfrac{\tau_{nd}}{n_e},$$

$$\nu_{ndl}^{(3)} = -\dfrac{5}{2}\dfrac{349}{64}\dfrac{k^2 T^2 n_e}{m_e}\dfrac{\tau_{nd}}{n_e} \quad (176)$$

$$= -\dfrac{1745}{128}\dfrac{k^2 T^2}{m_e}\tau_{nd} = -13.6328\dfrac{k^2 T^2}{m_e}\tau_{nd}.$$

The absolute values of diffusion and the diffusion thermoeffect obtained by the method of successive approximations of polynomials should be compared with the exact solution $\nu_{nd}^l$ and $\eta_{nd}^l$ obtained by the Lorentz method (168) and (169) for nondegenerate electrons

$$\eta_{nd}^{(l)} = -\dfrac{32}{3\pi}\dfrac{kT}{m_e n_e}\tau_{nd} = -3.3954\dfrac{kT}{m_e n_e}\tau_{nd}, \quad (177)$$

$$\nu_{nd}^{(l)} = -\dfrac{128}{3\pi}\dfrac{k^2 T^2}{m_e}\tau_{nd} = -13.5816\dfrac{k^2 T^2}{m_e}\tau_{nd}. \quad (178)$$

It is clear that the two-polynomial solution underestimates the value of the diffusion coefficient by 4.28% and the diffusion thermoeffect by 12.53%; and the three-polynomial solution differs from the exact one by approximately 0.14% for diffusion and by 0.38% for the diffusion thermoeffect. The equations in the case of the three-polynomial approximation in the presence of a magnetic field were obtained from (112) with allowance for (49)–(54) and (174) in the form

$$\begin{cases} -\dfrac{3}{2}\dfrac{\tau_{nd}}{n_e} = -\dfrac{3}{2} i\omega\tau_{nd} d_0 + \dfrac{3}{2} d_0 + \dfrac{9}{4} d_1 + \dfrac{45}{16} d_2, \\ 0 = -\dfrac{15}{4} i\omega\tau_{nd} d_1 + \dfrac{9}{4} d_0 + \dfrac{39}{8} d_1 + \dfrac{207}{32} d_2, \\ 0 = -\dfrac{105}{16} i\omega\tau_{nd} d_2 + \dfrac{45}{16} d_0 + \dfrac{207}{32} d_1 + \dfrac{1299}{128} d_2. \end{cases} \quad (179)$$

The explicit solution of equations (179) for two- and three-polynomial approximations is determined by formulas (122)–(132) with a formally infinite value $Z$.

### 6.3. Partially Degenerate Electrons

For a plasma with arbitrarily degenerate electrons in a magnetic field, system (173) will be presented, by analogy with [20], as follows:

$$\begin{cases} -\dfrac{3}{2} = -\dfrac{3}{2} i\omega n_e d_0 + d_0 b_{00} + d_1 b_{01} + d_2 b_{02}, \\ 0 = -\dfrac{15}{4}\left(\dfrac{7 G_{7/2}}{2 G_{3/2}} - \dfrac{5 G_{5/2}^2}{2 G_{3/2}^2}\right) i\omega n_e d_1 \\ \quad + d_0 b_{10} + d_1 b_{11} + d_2 b_{12}, \\ 0 = -\dfrac{105}{16}\left(-\dfrac{35 G_{7/2}^2}{8 G_{3/2}^2} + \dfrac{49 G_{7/2}^2}{2 G_{5/2}^2}\dfrac{G_{7/2}}{G_{3/2}}\right. \\ \quad \left. -\dfrac{63 G_{9/2}}{2 G_{5/2}}\dfrac{G_{7/2}}{G_{3/2}} + \dfrac{99 G_{11/2}}{8 G_{3/2}}\right) i\omega n_e d_2 + d_0 b_{20} \\ \quad + d_1 b_{21} + d_2 b_{22}. \end{cases} \quad (180)$$

For partially degenerate electrons with a level of degeneracy $DL = \varepsilon_{fe}/kT = 1.011$ and in the absence of a magnetic field, system (180) is written as

$$\begin{cases} -1.5 = d_0 b_{00} + d_1 b_{01} + d_2 b_{02}, \\ 0 = d_0 b_{10} + d_1 b_{11} + d_2 b_{12}, \\ 0 = d_0 b_{20} + d_1 b_{21} + d_2 b_{22}. \end{cases} \quad (181)$$

Taking into account the values of the coefficients $b_{jk}$ obtained in [20], this system can be rewritten in the form

$$\begin{cases} -1.5\dfrac{\tau_{d0}}{n_e} = 1.5d_0 + 2.16d_1 + 2.588d_2, \\ 0 = 2.16d_0 + 5.162d_1 + 6.671d_2, \\ 0 = 2.588d_0 + 6.671d_1 + 11.038d_2. \end{cases} \quad (182)$$

The first two equations for $d_2 = 0$ determine the two-polynomial approximation, which, taking into account (64) and (62), gives the following result:

$$d_0 = -2.5161\dfrac{\tau_{d0}}{n_e}, \quad d_1 = 1.0528\dfrac{\tau_{d0}}{n_e},$$
$$\eta^{(2)}_{d0l} = -2.516\dfrac{kT}{m_e n_e}\tau_{d0}, \quad \nu^{(2)}_{d0l} = -9.853\dfrac{k^2 T^2}{m_e}\tau_{d0}. \quad (183)$$

In the three-polynomial approximation, we obtain solution (182) for $d_0$ and $d_1$, as well as the diffusion and diffusion thermoeffect coefficients, in the form

$$d_0 = -2.591\tau_{d0}, \quad d_1 = 1.3658\tau_{d0},$$
$$\eta^{(3)}_{d0l} = -2.591\dfrac{kT}{m_e n_e}\tau_{d0}, \quad \nu^{(3)}_{d0l} = -10.873\dfrac{k^2 T^2}{m_e}\tau_{d0}. \quad (184)$$

The coefficients obtained by the method of successive approximations by polynomials should be compared with the exact solution $\eta^l_{nd}$ and $\nu^l_{nd}$ obtained by the Lorentz method (170) and (171) for nondegenerate electrons:

$$\eta^{(l)}_{d0} = -0.744\dfrac{32}{3\pi}\dfrac{kT}{m_e n_e}\tau_{d0} = -2.526\dfrac{kT}{m_e n_e}\tau_{d0}, \quad (185)$$

$$\nu^{(l)}_{d0} = -0.744\dfrac{128}{3\pi}\dfrac{k^2 T^2}{m_e}\tau_{d0} = -10.105\dfrac{k^2 T^2}{m_e}\tau_{d0}. \quad (186)$$

It can be seen that the two-polynomial solution for diffusion differs from the exact one by 0.4%, and the three-polynomial solution by 2.6%. For the diffusion thermoeffect, the two-polynomial solution differs by 2.5% from the exact one, and the three-polynomial solution differs by about 8% from the exact one.

## 7. CONCLUSIONS

In this paper, we found the tensors of kinetic coefficients of diffusion, thermal diffusion, and diffusion thermoeffect for nondegenerate electrons in a non-quantizing magnetic field. The tensors are obtained from the solution of the Boltzmann kinetic equation by the classical Chapman–Enskog method using the expansion into the Sonine polynomials and taking into account two and three terms of the expansion. Electron–ion and electron–electron collisions are taken into account. The tensors are written for an arbitrary local direction of the magnetic field and the temperature gradient in the Cartesian coordinate system according to [11]. Our results in the two-polynomial case are consistent with the results of previous works [8–10, 12, 13]. An analytical solution in the trinomial approximation had not been obtained previously.

The value of the thermal conductivity coefficients obtained in Braginsky's papers [12, 13] in the two-polynomial approximation is twice as small as the corresponding values for the thermal conductivity obtained in [20]. This is due to the approach used in [12], which differs from the classical Chapman–Enskog method [1]. In [12], it was believed that half of the heat flux is hidden in the so-called heat force in such a way that the resulting heat flux in the accompanying reference frame is the same in both considerations. The following values of the coefficients of thermal conductivity and electrical conductivity are obtained along the lines of the magnetic field ($B = 0$), at $Z = 1$:

$$\lambda^{\text{Braginski}}_{\parallel} = 3.1616\dfrac{n_e k^2 T \tau_{nd}}{m_e},$$
$$\lambda^{\text{2-pol}}_{\parallel} = 6.45\dfrac{n_e k^2 T \tau_{nd}}{m_e}, \quad (187)$$
$$\lambda^{\text{3-pol}}_{\parallel} = 7.6133\dfrac{n_e k^2 T \tau_{nd}}{m_e}.$$

In [15], the coefficients obtained numerically were compared with the work of Braginsky:

$$\lambda^{\text{E-H}}_{\parallel} = 3.203\dfrac{n_e T k^2 \tau_{nd}}{m_e}. \quad (188)$$

When we take into account that half of the heat flux is hidden in the thermal force, then $\lambda^{2-\text{pol}}_{\parallel}$ and $\lambda^{3-\text{pol}}_{\parallel}$ from (187), we obtain

$$\lambda'^{\text{2-pol}}_{\parallel} = 3.225\dfrac{n_e k^2 T \tau_{nd}}{m_e},$$
$$\lambda'^{\text{3-pol}}_{\parallel} = 3.8066\dfrac{n_e k^2 T \tau_{nd}}{m_e}. \quad (189)$$

The values of the conductivity coefficients from [15] and from this paper, obtained in the two- and three-polynomial approximations, coincide against each other up to fractions of tenths:

$$\sigma^{\text{E-H}}_{\parallel} = 1.9763\dfrac{n_e^2 \tau_{nd}}{m_e}, \quad \sigma^{\text{2-pol}}_{\parallel} = 1.9319\dfrac{n_e^2 \tau_{nd}}{m_e}, \quad (190)$$
$$\sigma^{\text{3-pol}}_{\parallel} = 1.9497\dfrac{n_e^2 \tau_{nd}}{m_e}.$$

Using the Lorentz approximation as an example, it was shown that the accuracy of approximation by a number of orthogonal functions similar to the Sonine polynomials decreases with an increasing degree of degeneracy. For nondegenerate electrons, excluding

FOUR TENSORS DETERMINING THE THERMAL AND ELECTRIC CONDUCTIVITIES

electron–electron collisions, the value of the thermal diffusion coefficient in the two-polynomial approximation underestimates the exact solution obtained in the Lorentz approximation at $B = 0$ by 26%, and in the three-polynomial approximation it overestimates it by approximately 1%. For partially degenerate electrons, with $x_0 = 0$, the two-polynomial solution underestimates the exact one by 28%, and the three-polynomial solution underestimates the exact one by 7%. It should be noted that electron–electron collisions further reduce the coefficient.

The diffusion coefficient obtained from the two-polynomial approximation underestimates the exact solution by about 4%, and the three-polynomial solution underestimates the exact solution by 0.14% for nondegenerate electrons. In the case of partial degeneracy, the diffusion obtained from the two-polynomial solution underestimates the exact one by 0.4%, and the diffusion obtained from the three-polynomial solution overestimates the exact one by 3%.

The coefficient of the diffusion thermoeffect for nondegenerate electrons in the two-polynomial approximation is underestimated by 12%, and in the three-polynomial approximation, it is overestimated by 0.4% in comparison to the exact solution. With the partial degeneration of electrons, the two-polynomial solution underestimates the exact solution by 2.5%, and the three-polynomial solution overestimates the exact solution by 8%.

The Chapman–Enskog method can be used for a sufficiently dense gas (plasma), where the time between particle collisions is the smallest value among other characteristic times. In the presence of a magnetic field, in addition to the lifetime of the system and the characteristic time of changes in the parameters in the plasma, the rotation time along the Larmor circle $\tau_L = 2\pi/\omega$ is added. This time should be much shorter than $\tau$, of the order of $\tau_{nd}$, which leads to the condition that you can use the Chapman–Enskog method, in the form

$$\omega\tau \ll 2\pi. \tag{191}$$

Therefore, the results of this paper can be successfully applied for cases when $\omega\tau \lesssim 1$, and for the case with large $\omega\tau$ only qualitative estimates can be obtained.

ACKNOWLEDGMENTS

The author thanks G.S. Bisnovatyi-Kogan for his discussions and help.

FUNDING

This work was supported by the Russian Science Foundation (project no. 18-12-00378).